\documentclass[aps,prd, groupedaddress,amssymb,preprint,showpacs,epsfig,nofootinbib]{revtex4}

\usepackage{graphicx}
\usepackage{bm}
\usepackage{dcolumn}

\usepackage{amsmath}
\usepackage{amssymb}

\usepackage{epsfig}

\usepackage{graphicx}
\usepackage{bm}
\usepackage{dcolumn}

\usepackage{amsmath}
\usepackage{amssymb}

\usepackage{epsfig}

\begin{document}
\newcommand{\newc}{\newcommand}

\newc{\be}{\begin{equation}}
\newc{\ee}{\end{equation}}
\newc{\ba}{\begin{eqnarray}}
\newc{\ea}{\end{eqnarray}}
\newc{\bea}{\begin{eqnarray*}}
\newc{\eea}{\end{eqnarray*}}
\newc{\D}{\partial}
\newc{\ie}{{\it i.e.} }
\newc{\eg}{{\it e.g.} }
\newc{\etc}{{\it etc.} }
\newc{\etal}{{\it et al.}}
\newcommand{\nn}{\nonumber}
\newc{\ra}{\rightarrow}
\newc{\lra}{\leftrightarrow}
\newc{\lsim}{\buildrel{<}\over{\sim}}
\newc{\gsim}{\buildrel{>}\over{\sim}}
\newcommand{\mincir}{\raise
-3.truept\hbox{\rlap{\hbox{$\sim$}}\raise4.truept\hbox{$<$}\ }}
\newcommand{\magcir}{\raise
-3.truept\hbox{\rlap{\hbox{$\sim$}}\raise4.truept\hbox{$>$}\ }}
\def\ft#1#2{{\textstyle{\frac{\scriptstyle #1}{\scriptstyle #2} } }}
\def\fft#1#2{{\frac{#1}{#2}}}
\def\del{\partial}
\def\vp{\varphi}
\def\sst#1{{\scriptscriptstyle #1}}
\def\oneone{\rlap 1\mkern4mu{\rm l}}
\def\td{\tilde}
\def\wtd{\widetilde}
\def\ie{{\it i.e.\ }}
\def\dalemb#1#2{{\vbox{\hrule height .#2pt
        \hbox{\vrule width.#2pt height#1pt \kern#1pt
                \vrule width.#2pt}
        \hrule height.#2pt}}}
\def\square{\mathord{\dalemb{6.8}{7}\hbox{\hskip1pt}}}
\newcommand{\pd}{\partial}
\newcommand{\ud}{\textrm{d}}
\newcommand{\dTH}{T^{\prime \, 0}_\textrm{H}}
\newcommand{\dOi}{\Omega^{\prime \, 0}_i}
\newcommand{\bx}{{\bf x}}
\def\0{{\sst{(0)}}}
\def\1{{\sst{(1)}}}
\def\2{{\sst{(2)}}}
\def\3{{\sst{(3)}}}
\def\4{{\sst{(4)}}}
\def\5{{\sst{(5)}}}
\def\6{{\sst{(6)}}}
\def\7{{\sst{(7)}}}
\def\8{{\sst{(8)}}}
\def\m{{\sst{(m)}}}
\def\n{{\sst{(n)}}}
\def\cA{{{\cal A}}}
\def\cB{{{\cal B}}}
\def\cF{{{\cal F}}}
\def\cG{{{\cal G}}}
\def\cH{{{\cal H}}}
\def\tV{\widetilde V}
\def\tW{\widetilde W}
\def\tH{\widetilde H}
\def\tE{\widetilde E}
\def\tF{\widetilde F}
\def\tA{\widetilde A}
\def\im{{{\rm i}}}

\title{The absence of the  Kerr black hole \\in the Ho\v{r}ava-Lifshitz gravity}

\author{Hyung Won Lee}
\email{hwlee@inje.ac.kr}

\author{Yun Soo Myung}
\email{ysmyung@inje.ac.kr}

\affiliation{Institute of Basic Science and School of Computer Aided
Science, Inje University, Gimhae 621-749, Korea}

\begin{abstract}
We show that the Kerr metric does not exist as  a fully rotating
black hole solution to the modified Ho\v{r}ava-Lifshitz (HL) gravity
with $\Lambda_W=0$ and $\lambda=1$ case. We perform it by showing
that the Kerr metric does not satisfy  full equations derived from
the modified HL gravity.
\end{abstract}

\maketitle

\section{Introduction}
Ho\v{r}ava has proposed a renormalizable theory of modified gravity
at a Lifshitz point~\cite{ho1,ho2},  which  may be regarded as a UV
complete candidate for general relativity. At short distance of the
UV scale, the Ho\v{r}ava-Lifshitz (HL) gravity describes interacting
non-relativistic gravitons and is supposed to be power counting
renormalizable in four dimensions.  However, we would like to stress
that it belongs to a Lorentz-violating gravity theory even though
the Lorentz-symmetry is
 expected  hopefully  to be recovered in the IR limit.

On the other hand, its black hole solutions  has been intensively
investigated in
~\cite{LMP,CCO1,CLS,CY,MK,KS,CCO2,Gho,Myung,CJ1,park,Gho09,lkme,Myungent,KKBH,lifMyung}.
Concerning spherically symmetric solutions, L\"u-Mei-Pope (LMP) have
obtained the black hole solution with dynamical parameter
$\lambda$~\cite{LMP} and topological black holes were found in
\cite{CCO1}. Its thermodynamics were studied in \cite{CCO2}, but
there remain unclear issues in obtaining the ADM mass and entropy
because their asymptotic spacetimes is Lifshitz~\cite{MK}. On the
other hand, Kehagias and Sfetsos (KS) have found the $\lambda=1$
black hole solution in asymptotically flat spacetimes considering
the modified HL gravity~\cite{KS}. Its thermodynamics was
investigated in Ref.\cite{Myung,Myungent}. Park has obtained a
$\lambda=1$ black hole solution with $\omega$ and
$\Lambda_W$~\cite{park}.

It is very interesting to find a fully rotating black hole solution
in the HL gravity since the HL gravity is being considered as a
promising modified gravity which violates the Lorentz-symmetry.
However, it is a formidable task to find a fully rotating solution
because equations of motion to be solved are very complicated.
Fortunately, slowly rotating black holes  based on the
KS~\cite{LKMslowly} and LMP~\cite{AS} solutions were found in the HL
gravity. Here ``slowly rotating black hole" means that one considers
up to linear order of rotation parameter $a=J/M$ in the metric
functions, equations of motion, and thermodynamic quantities. This
implies that the case of $a\ll1$ is valid for slowly rotating black
hole.  We  mention that the slowly rotating Kerr
 black hole could be  recovered from the slowly
rotating KS black hole solutions in the limit of $\omega \to
\infty$. In this case, the role of $\omega$ is neglected and thus,
the HL gravity reduces to general relativity.

The above case is similar to the parity-violating Chern-Simons (CS)
modified gravity~\cite{JP}. Since the parity-violation may imply the
breaking of Lorentz symmetry, the CS modified gravity may belong to
the Lorentz-violating theory.   It is well known that the CS term
could not yield the Kerr metric as a fully rotating black hole
solution since the Pontryagin constraint ${}^*RR=0$ required by the
Bianchi identity is not satisfied.  Up to now, the slowly rotating
Kerr black hole is known to be  the only solution to the  CS
modified gravity~\cite{sCS}.

Recently, the Penrose process on rotational energy extraction of the
black hole in the HL gravity was studied by considering that the
Kerr solution is a truly rotating solution to the HL
gravity~\cite{AAA}. This approach might be incorrect because the
Kerr solution is not yet proved to be a fully  rotating solution to
the modified HL gravity.

In this work, we wish to  show that the Kerr metric does not exist
as a fully rotating solution to  the modified HL gravity because
Lorentz-violating higher order terms are present. Our strategy is to
prove the non-existence  by plugging the Kerr metric directly to
full equations derived from the modified HL gravity. In achieving
the non-existence of the Kerr black hole, the order of rotation
parameter $a$ plays an important role in classifying full equations.
This implies that three full equations are classified according to
the order of $a$ and then,  we check whether or not each equation in
$a$-order is satisfied. This approach is inspired by finding the
slowly rotataing black hole.

\section{HL gravity}
In this section, we review briefly the modified HL gravity including
a soft-violation term. The ADM formalism implies that the
four-dimensional metric of general relativity is parameterized as
\be
ds_4^2= - N^2  dt^2 + g_{ij} (dx^i - N^i dt) (dx^j - N^j dt)\,,
\label{metricans}
\ee
where the lapse $N$, shift  $N^i$,  and three-dimensional space
metric $g_{ij}$ are all functions of $t$ and $x^i$. The ADM
decomposition  of the Einstein-Hilbert action with a cosmological
constant $\Lambda$ is given by
\be S_{EH} = \fft{1}{16\pi G} \int d^4x \sqrt{g} N \Big(K_{ij}
K^{ij} - K^2 + R - 2\Lambda \Big)\,,\label{ehlag} \ee
where $G$ is Newton's constant and extrinsic curvature $K_{ij}$ is
defined by
\be K_{ij} = \fft{1}{2N} \Big(\dot g_{ij} - \nabla_i N_j - \nabla_j
N_i\Big)\,. \ee
Here a over-dot denotes a derivative with respect to $t$.

The action of the modified HL theory  including a soft-violation
term $\mu^4 R$ is given by  \be S=\int dtd^3 \bx\, \Big({\cal L}_0 +
{\cal L}_1\Big), \label{hlaction0}\ee
 \be {\cal L}_0= \sqrt{g}N\Big\{\frac{2}{\kappa^2}(K_{ij}K^{ij} -\lambda
K^2)+\frac{\kappa^2\mu^2(\Lambda_W R
  -3\Lambda_W^2)}{8(1-3\lambda)} + \mu^4R \Big\},\label{hlaction1}\ee

 \be {\cal L}_1=
\sqrt{g}N\Big\{\frac{\kappa^2\mu^2 (1-4\lambda)}{32(1-3\lambda)}R^2
-\frac{\kappa^2}{2W^4} \Big(C_{ij} -\frac{\mu W^2}{2}R_{ij}\Big)
\Big(C^{ij} -\frac{\mu W^2}{2}R^{ij}\Big)\Big\},\label{hlaction2}\ee
where $\lambda\,,\kappa\,,\mu\,,$ and $W$  are constant parameters
to represent  the modified HL gravity and  $\Lambda_W$ is a negative
cosmological constant.   Here, $\mu^4$ can be expressed in terms of
$\omega$ as \be \mu^4=\frac{\kappa^2\mu^2
\omega}{8(3\lambda-1)},~~\omega=\frac{8(3\lambda-1)\mu^2}{\kappa^2}
\ee and
 $C_{ij}$ is the Cotton tensor defined by
\be C^{ij}=\epsilon^{ik\ell}\nabla_k\left(R^j{}_\ell
-\frac14R\delta_\ell^j\right) \,.\label{cotton} \ee
Since we wish to find a non-spherical solution of the black hole, we
have to know full equations of motion to the action
(\ref{hlaction0}), which are composed of three equations. The
equation from variation of the lapse function $N$ is given by
\be \fft{2}{\kappa^2}(K_{ij}K^{ij} - K^2) +\frac{\kappa^2\mu^2 \left
\{(\Lambda_W - \omega) R -3\Lambda_W^2 \right \}}{8(1-3\lambda)}
+\frac{\kappa^2\mu^2 (1-4\lambda)}{32(1-3\lambda)}R^2 -
\frac{\kappa^2}{2W^4} Z_{ij} Z^{ij}=0\,\label{eom1}
\ee%
with
\be Z_{ij}\equiv C_{ij} - \fft{\mu W^2}{2} R_{ij}\,. \ee
We will focus on this lapse equation (\ref{eom1}) for testing the
Kerr metric as the solution to the HL gravity.  The variation
$\delta N^i$ implies an equation
\be
\nabla_k(K^{k\ell}-\lambda\,Kg^{k\ell})=0\,.\label{eom2}
\ee%
Equation of motion  from variation of $\delta g^{ij}$ is complicated
to be~\cite{LMP}
\be%
\frac{2}{\kappa^2}E_{ij}^\1-\frac{2\lambda}{\kappa^2}E_{ij}^\2
+\frac{\kappa^2\mu^2(\Lambda_W-\omega)}{8(1-3\lambda)}E_{ij}^\3
+\frac{\kappa^2\mu^2(1-4\lambda)}{32(1-3\lambda)}E_{ij}^\4
-\frac{\mu\kappa^2}{4W^2}E_{ij}^\5
-\frac{\kappa^2}{2W^4}E_{ij}^\6=0, \label{eom3}\ee
where \bea
E_{ij}^\1&=& N_i \nabla_k K^k{}_j + N_j\nabla_k K^k{}_i -K^k{}_i
\nabla_j N_k-
   K^k{}_j\nabla_i N_k - N^k\nabla_k K_{ij}\nn\\
&& - 2N K_{ik} K_j{}^k
  -\frac12 N K^{k\ell} K_{k\ell}\, g_{ij} + N K K_{ij} + \dot K_{ij}
\,,\nn \\
E_{ij}^\2&=& \frac12 NK^2 g_{ij}+ N_i \pd_j K+
N_j \pd_i K- N^k (\pd_k K)g_{ij}+  \dot K\, g_{ij}\,,\nn\\
E_{ij}^\3&=&N\Big(R_{ij}- \frac12Rg_{ij}+\frac32
\frac{\Lambda_W^2}{\Lambda_W-\omega} g_{ij}\Big)-(
\nabla_i\nabla_j-g_{ij}\nabla_k\nabla^k)N\,,\nn\\
E_{ij}^\4&=&NR\Big(2R_{ij}-\frac12Rg_{ij}\Big)- 2
\big(\nabla_i\nabla_j
-g_{ij}\nabla_k\nabla^k\big)(NR)\,,\nn\\
E_{ij}^\5&=&\nabla_k\big[\nabla_j(N Z^k_{~~i}) +\nabla_i(N
Z^k_{~~j})\big]  -\nabla_k\nabla^k(NZ_{ij})
-\nabla_k\nabla_\ell(NZ^{k\ell})g_{ij}\,,\ \nn 
\eea

\bea \label{eeeq} E_{ij}^\6&=&-\frac12NZ_{k\ell}Z^{k\ell}g_{ij}+
2NZ_{ik}Z_j^{~k}-N(Z_{ik}C_j^{~k}+Z_{jk}C_i^{~k})
+NZ_{k\ell}C^{k\ell}g_{ij}\nn\\
&&-\frac12\nabla_k\big[N\epsilon^{mk\ell}
(Z_{mi}R_{j\ell}+Z_{mj}R_{i\ell})\big]
+\frac12R^n{}_\ell\, \nabla_n\big[N\epsilon^{mk\ell}(Z_{mi}g_{kj}
+Z_{mj}g_{ki})\big]\nn\\
&&-\frac12\nabla_n\big[NZ_m^{~n}\epsilon^{mk\ell}
(g_{ki}R_{j\ell}+g_{kj}R_{i\ell})\big]
-\frac12\nabla_n\nabla^n\nabla_k\big[N\epsilon^{mk\ell}
(Z_{mi}g_{j\ell}+Z_{mj}g_{i\ell})\big]\nn\\
&&+\frac12\nabla_n\big[\nabla_i\nabla_k(NZ_m^{~n}\epsilon^{mk\ell})
g_{j\ell}+\nabla_j\nabla_k(NZ_m^{~n}\epsilon^{mk\ell})
g_{i\ell}\big]\nn\\
&&+\frac12\nabla_\ell\big[\nabla_i\nabla_k(NZ_{mj}
\epsilon^{mk\ell})+\nabla_j\nabla_k(NZ_{mi}
\epsilon^{mk\ell})\big] \nn \\
&&-\nabla_n\nabla_\ell\nabla_k
(NZ_m^{~n}\epsilon^{mk\ell})g_{ij}\,.\eea%
We note that in deriving these equations, we have relaxed both the
projectability  and detailed-balance conditions since the lapse
function $N$  depends on the spatial coordinate $x^i$ as well as the
soft-violation term of $\mu^4 R$ is included.

Hereafter, we will focus on  the case of $\Lambda_W=0,~\lambda=1$
and $\omega=16\mu^2/\kappa^2$, providing asymptotically flat
spacetimes. In this case, we may have the  Minkowski background as
vacuum solution with ~\cite{KS} \be
c^2=\fft{\kappa^2\mu^4}{2}\,,\qquad
G=\fft{\kappa^2}{32\pi\,c}\,,\qquad \Lambda=0\,.\label{mcg} \ee

\section{Kehagias-Sfetsos and its slowly rotating solutions}
In this section, we  investigate  how the KS and its slowly rotating
black holes are derived  in asymptotically flat spacetimes. First of
all, we would like to find  a static solution, the KS solution by
considering a spherically symmetric  line element
\begin{equation}
ds^2 = -N(r)^2 dt^2 + \frac{dr^2}{f(r)} + r^2 \left ( d \theta^2 +
\sin^2 \theta d \phi^2 \right ) . \label{sph_ansatz}
\end{equation}
In this case, we find that $K_{ij}=0$ and $C_{ij}=0$. Equation for
the lapse function $N$ can be read as
\be \mu^4 \Big[ R -\frac{2 }{\omega}\Big(R_{ij}^2 - \frac{3}{8}
R^2\Big)\Big]=0\,.\label{eom1-KS}
\ee%
 This leads to the first-order
equation for $f$ as
\begin{equation}
\frac{(f-1)^2}{r^2}-\frac{2(f-1)f'}{r}-2\omega(1-f-rf') =0.
\label{eom1_KS 0}
\end{equation}
The equation (\ref{eom2}) from  the shift function $\delta N^i$ is
trivially satisfied. The equation  from $\delta g^{ij}$ reduces to
\be \frac{\mu^2 \kappa^2}{16} \left[ \omega E_{ij}^\3
+\frac{3}{4}E_{ij}^\4\right] - \frac{\mu\kappa^2}{4W^2} \left [
E_{ij}^\5 + \frac{2}{\mu W^2}E_{ij}^\6 \right ] =0.\label{eom3-KS}
\ee The $(rr)$-component of the above equation becomes
\begin{equation}
-\frac{\mu^2 \kappa^2}{32} \frac{1}{r^4 f} \Big[ N' 4 r f \left ( -1
+f - \omega r^2 \right )
 - N \left ( 1 + f^2 - 2\omega r^2 -2 f + 2\omega r^2 f \right ) \Big] = 0,
\label{delg_rr}
\end{equation}
which leads to
\begin{equation}
\frac{N'}{N} = \frac{1 + f^2 - 2 \omega r^2 -2 f + 2 \omega r^2 f}
{4 r f \left ( -1 + f - \omega r^2 \right ) } .
\label{delg_rr1}
\end{equation}
On the other hand, from (\ref{eom1_KS 0}) we have   $f'$ as
\begin{equation}
f' = \frac{1 + f^2 - 2 \omega r^2 -2 f + 2 \omega r^2 f}
{2 r \left ( -1 + f - \omega r^2 \right ) }.
\label{sol_fp}
\end{equation}
Substituting this  into (\ref{delg_rr1}), we obtain
\begin{equation}
\frac{N'}{N} = \frac{1}{2} \frac{f'}{f},
\label{eq_log}
\end{equation}
which implies
\begin{equation}
N^2=f. \label{eom1_KS 1}
\end{equation}
Hence, the KS solution to Eq. (\ref{eom1_KS 0})  is given by
\begin{equation}
f_{\rm KS} =N^2_{\rm KS}= 1 + \omega r^2 \left ( 1 - \sqrt{1 +
\frac{4 M}{\omega r^3}}\right ). \label{sph-sol}
\end{equation}
 In the limit of $\omega \to
\infty$ (equivalently, the decoupling limit of higher curvature
terms), it reduces to the Schwarzschild form of \be
 f_{\rm Sch}(r)=1-\frac{2M}{r}. \ee

Now we consider an axisymmetric metric  for finding a slowly
rotating KS black hole
\begin{equation}
ds^2_{\rm sr} = -f(r) dt^2 + \frac{dr^2}{f(r)} + r^2 d \theta^2 +
       r^2 \sin^2 \theta \Big[ d \phi^2 - 2a N^\phi(r) dt d\phi \Big].
\label{axisr_ansatz}
\end{equation}
Then, the  extrinsic curvature tensor  is found to be
\begin{equation}
K_{ij} = \begin{pmatrix}
         0 & 0 & -\frac{1}{2} \frac{r^2 a \sin^2 \theta}{\sqrt{f(r)}} \frac{dN^\phi(r)}{dr} \\
         0 & 0 &0 & \\
         -\frac{1}{2} \frac{r^2 a \sin^2 \theta}{\sqrt{f(r)}} \frac{dN^\phi(r)}{dr} & 0 & 0
        \end{pmatrix},
\label{K-ij}
\end{equation}
where $K_{r\phi}=K_{\phi r}\not=0$ are  linear order of $a$, but
$K=0$.  We note here that all components of Cotton tensor still
vanish ($C_{ij}=0$), since it is constructed  from the metric
$g_{ij}$ which does not include the rotation parameter $a$. This
implies that if one wishes to find a fully rotating black hole, all
higher order terms of $a$ must be included.  Using (\ref{K-ij}), Eq.
(\ref{eom2}) reduces to \be \nabla_k K^{k\ell}=0 \to {\rm diag}
\left [ 0, 0, \frac{a \sqrt{f(r)}}{2r}
      \left (  r \frac{d^2 N^\phi(r)}{dr^2} + 4 \frac{dN^\phi(r)}{dr} \right ) \right ] = 0\,,\label{eom2_KS_reduce}
\ee%
which has a solution with two unknown constants $C_1$ and $C_2$
\begin{equation}
N^\phi(r) = C_1 + \frac{C_2}{r^3}.
\label{N3_sol}
\end{equation}
For later convenience, we choose the shift function to be \be
N^\phi(r) = \frac{2M}{r^3} \ee with $C_1=0$ and $C_2=2M$. In this
case, one has the  $g_{t\phi}$-component
\begin{equation}
g_{t\phi}=-ar^2 N^\phi(r)\sin^2\theta=-\frac{2J}{r}\sin^2\theta.
\end{equation}
Plugging these into Eq.(\ref{eom1}) leads to  \be \mu^4 \Big[ R
+\frac{2 }{\omega}\Big(R_{ij}^2 - \frac{3}{8}
R^2\Big)\Big]=-\frac{2}{\kappa^2}\Big(K_{ij}K^{ij}-K^2\Big)\label{eom1-KS-1}
\ee%
which takes the form
\begin{equation}
\frac{(f-1)^2}{r^2}-\frac{2(f-1)f'}{r}-2\omega(1-f-rf') =
\frac{32a^2 \omega M^2 \sin^2 \theta}{\kappa^4\mu^4r^4} \simeq 0.
\label{eom1_KS_reduce}
\end{equation}
Here, we  take the right-hand side  to be zero effectively because
it is second order of $a$.   Then, the solution is given by the
KS-type in Eq.(\ref{sph-sol}).

Consequently, the slowly  rotating  KS black hole solution is given
by
\begin{equation}
ds^2_{\rm srKS} = -f_{\rm KS}(r) dt^2 + \frac{dr^2}{f_{\rm KS}(r)} +
r^2 d \theta^2 +
       r^2 \sin^2 \theta \left ( d \phi^2 - \frac{4J}{r^3} dt d\phi \right ).
\label{slowrsol}
\end{equation}
In the limit of $\omega \to \infty$, it leads to the slowly rotating
Kerr black hole as
\begin{equation}
ds^2_{\rm srKerr} = -f_{\rm Sch}(r) dt^2 + \frac{dr^2}{f_{\rm
Sch}(r)} + r^2 d \theta^2 +
       r^2 \sin^2 \theta \left ( d \phi^2 - \frac{4J}{r^3} dt d\phi \right ).
\label{srKerr}
\end{equation}
In this sense, we wish to clarify that the slowly rotating Kerr
black hole is not the solution to the HL gravity but the Einstein
gravity, on the contrary to Ref.~\cite{SP}.

Finally, we explain why the slowly rotating solution is naturally
obtained for the HL gravity by examining the order of rotation
parameter $a$ in equations of motion. The axisymmetric metric ansatz
(\ref{axisr_ansatz}) was implemented by one-component shift vector
of $N^\phi$.  Hence, extrinsic curvature $K_{ij}$ has off-diagonal
components as shown in (\ref{K-ij}).  This implies that equation
(\ref{eom1-KS-1}) obtained from   $\delta N$ remains unchanged when
adding a rotating parameter term   to the spherically symmetric
case.  This is confirmed by showing that $K=0$ and $K_{ij}=0$
identically for spherically symmetric case, while  $K=0$ but $K_{ij}
K^{ij} = {\cal{O}}(a^2)$ for slowly rotating case.  Effectively, Eq.
(\ref{eom1-KS-1}) is the same equation  for both two cases.  A shift
vector $N^{\phi}$  could be determined by Eq.
(\ref{eom2_KS_reduce}). We emphasize that Eq. (\ref{eom3}) remains
unchanged at linear order of $a$. It is clear that $E^{(1)}_{ij}=0$
for spherically symmetric case and $E^{(1)}_{ij}={\cal{O}}(a^2)$ for
slowly rotating case, which is effectively taken to be zero.
$E^{(2)}_{ij}=0$ for  both two cases. All other $E^{(r)}_{ij}$ for
$r = 3, \cdots, 6$ remain unchanged, since they contain  terms
derived from $g_{ij}$ which does not carry $a$, and thus, $C_{ij} =
0$ by rotation symmetry in three-dimensional Euclidean space.

\section{Kerr metric is not a rotating solution to the HL gravity}
In this section, we wish  to check explicitly whether or not  the
Kerr metric is a solution to the  HL gravity.  For this purpose, we
introduce the Kerr line-element written in Boyer-Lindquist
coordinates as
\begin{equation}
ds^2_{\rm Kerr} = - \frac{\rho^2 \Delta_r}{\Sigma^2} dt^2 +
\frac{\rho^2}{\Delta_r} dr^2
     + \rho^2 d\theta^2 + \frac{\Sigma^2 \sin^2 \theta}{\rho^2}
       \left ( d\phi - \xi dt \right )^2,
\label{kerr-metric}
\end{equation}
where
\begin{eqnarray}
\rho^2 &=& r^2 + a^2 \cos^2 \theta, \nonumber \\
\Delta_r &=& \left ( r^2 + a^2 \right ) - 2 M r, \nonumber \\
\Sigma^2 &=& \left ( r^2 + a^2 \right )^2 - a^2 \sin^2 \theta \Delta_r, \nonumber \\
\xi &=& \frac{2Mar}{\Sigma^2}.
\end{eqnarray}
From this metric we identify the lapse $N^2$, shift $N^\phi$ and
three-dimensional metric $g_{ij}$  with, respectively,
\begin{equation}
N^2(r, \theta) = \frac{\rho^2 \Delta_r}{\Sigma^2}, \,\,
N^{\phi} = \xi = \frac{2Mar}{\Sigma^2},\,\,
g_{rr} = \frac{\rho^2}{\Delta_r}, \,\,
g_{\theta\theta} = \rho^2, \,\,
g_{\phi\phi} = \frac{\Sigma^2 \sin^2\theta}{\rho^2}.
\end{equation}
In the limit of $a\to 0$, the Kerr spacetimes (\ref{kerr-metric})
reduces to the Schwarzschild spacetimes: (\ref{srKerr}) with $J=0$.

 Using
Eq.(\ref{kerr-metric}), the extrinsic curvature tensor is computed
to be
\begin{equation}
K_{ij} = \begin{pmatrix}
         0 & 0 & K_{r\phi} \\
         0 & 0 & K_{\theta\phi} \\
         K_{r\phi} & K_{\theta\phi} & 0
        \end{pmatrix},
\label{K_ij_kerr}
\end{equation}
where
\begin{eqnarray}
K_{r\phi} &=&
\frac{Ma \sin^2 \theta \left [ 3r^4 + r^2 a^2 + a^2 \left ( r^2 + a^2 \right ) \cos^2 \theta \right ] }
{K_d} K_f, \\
K_{\theta\phi} &=&
-\frac{2\, Mr{a}^{3} \sin^3 \theta \cos \theta  \left( r^2 + {a}^{2}-2\,Mr \right) }
{K_d} K_f,
\end{eqnarray}
with
\begin{eqnarray}
K_d &=& {a}^{4}\cos^4 \theta \left ( r^2 + a^2 - 2Mr \right )
  + 2 r a^2 \cos^2 \theta \left ( r^3 - M r^2 + r a^2 + M a^2 \right ) \nonumber \\
&&
  + r^3 \left ( r^3 + r a^2 + 2M a^2 \right ) , \\
K_f^2 &=&  \frac {{r}^{4}+{r}^{2}{a}^{2}+2\,Mr{a}^{2}\sin^2 \theta +
a^2 \left ( r^2 + a^2 \right ) \cos^2 \theta } { \left(
{r}^{2}+{a}^{2} \cos^2 \theta \right) \left( r^2 + {a}^{2}-2\,Mr
\right) }.
\end{eqnarray}
The  Ricci tensor takes the form
\begin{equation} R_{ij}=\begin{pmatrix}
         R_{rr} & R_{r\theta} & 0 \\
         R_{r\theta} & R_{\theta\theta} & 0 \\
         0 &0 & R_{\phi\phi}
        \end{pmatrix}.
\label{R_ij_kerr}
\end{equation}
whose explicit form  is  given in Appendix \ref{apendixA}. The
lowest order of $R_{ij}$ in $a$ is
\begin{eqnarray}
R^{(0)}_{rr} &=&
\frac{2M}{r^3}\frac{1}{1-\frac{2M}{r}},~~R^{(0)}_{\theta\theta}=-\frac{M}{r},
~~R^{(0)}_{\phi\phi}=-\frac{M}{r}\sin^2\theta,
\nn \\
 R^{(2)}_{r\theta} &=&-\frac{9M a^2 \sin \theta \cos \theta }{ r^4}.
\end{eqnarray}
The Ricci scalar is given by
\begin{equation}
R^{(2)}=-\frac{18M^2a^2\sin^2\theta}{r^6}.
\end{equation}
The Cotton tensor has
\begin{equation} C_{ij}=\begin{pmatrix}
         0 & 0 & C_{r\phi} \\
         0 & 0 & C_{\theta\phi} \\
         C_{r\phi} &C_{\theta\phi} & 0
        \end{pmatrix},
\label{C_ij_kerr}
\end{equation}
where
\begin{eqnarray}
C_{r\phi} &=& -2\,{M}^{2}{a}^{2}\cos \theta F_{r\phi} \frac{N_{r\phi}}{D_{r\phi}}, \label{c12}\\
C_{\theta\phi} &=& 2\, M^2 a^2 \sin \theta F_{r\phi} \frac{N_{\theta\phi}}{D_{\theta\phi}}. \label{c23}
\end{eqnarray}
The explicit forms of $F_{r\phi}$, $N_{r\phi}$, $D_{r\phi}$,
$N_{\theta\phi}$, and $D_{\theta\phi}$  are given in Appendix
\ref{apendixB}. The lowest-order term of $C_{ij}$ in $a$ takes the
form
\begin{eqnarray}
C^{(2)}_{r\phi} = -\frac{2{M}^{2}{a}^{2}\cos \theta}{r^8},~~
C^{(2)}_{\theta\phi} = -\frac{2M^2 a^2 \sin \theta}{ r^9}. \label{c023}
\end{eqnarray}

 The Kerr metric satisfies Eq.(\ref{eom2}) with
$\lambda=1$ and $\Lambda_{\rm W}=0$ because it does not contain any
HL gravity parameters, as in the Einstein gravity.

 However, the Kerr
metric does not satisfy Eq. (\ref{eom1}) since there exist
higher-order curvature terms.  In order to see it explicitly, we
rewrite Eq. (\ref{eom1}) as
\begin{equation}
 \frac{2}{\kappa^2} \left ( K_{ij}K^{ij} - K^2  + R \right )
+\left ( \mu^4 - \frac{2}{\kappa^2} \right )  R
+\frac{3\kappa^2\mu^2 }{64}R^2 - \frac{\kappa^2}{2W^4} Z_{ij} Z^{ij}
=0. \label{eom1-split}
\end{equation}
We note that the first term is  from the Einstein-Hilbert action and
thus, it vanishes for the Kerr metric.  The second term is zero when
choosing $c^2=1(\mu^4=2/\kappa^2)$. The last two higher-order terms
survive as
\begin{equation}\label{lowesteq0}
-\frac{\kappa^3}{2^{7/2}}\left[R^2_{ij}-\frac{3}{8}R^2\right]
-\frac{\kappa^4}{4W^4}C_{ij}^2=0,
\end{equation}
because $R\not=0$, $R_{ij}\not=0$, $C_{ij}\not=0$, but
$C_{ij}R^{ij}=0$ for the Kerr solution. In order for
(\ref{lowesteq0}) to be satisfied, each term should vanish  because
each term has different power of $\kappa$ and $W$, being considered
as independent parameters. Using $\kappa^3=2^{9/2}/\omega$, $\omega$
and $W$ are also  regarded as two independent parameters.  The
explicit form of $R^2$, $R_{ij}^2$, and $C_{ij}^2$ are given in
Appendix \ref{apendixC} upto $a^4$-order.  Eq. (\ref{lowesteq0}) is
split  into according to  the order of $a$
\begin{itemize}
\item $a^0$-order:
\begin{equation} \label{zerotheq}
-\frac{\kappa^3}{2^{7/2}} \left [\frac{6M^2}{r^6} \right ]
-\frac{\kappa^4}{4W^4} \Big[0 \Big] \ne 0,
\end{equation}
\item $a^2$-order:
\begin{equation}
-\frac{\kappa^3}{2^{7/2}} \left [ \frac{27 M^2 \sin^2
\theta}{4r^6} +\frac{18M^2}{r^8} \left ( 1 -5 \cos^2 \theta -
\frac{3M\sin \theta}{r}\right ) \right ] -\frac{\kappa^4}{4W^4}
\Big[0 \Big] \ne 0,
\end{equation}
\item $a^4$-order:
\begin{eqnarray}
&&-\frac{\kappa^3}{2^{7/2}} \left [
-\frac{9M^2 \sin^2 \theta}{4r^8}
 \left ( 4 + 13 \cos^2 \theta + \frac{12 M \sin^2 \theta}{r}\right ) \right . \nonumber \\
&& \left .
+\frac{6M^2}{r^{10}} \left \{
\frac{81 M^2 \sin^4 \theta}{r^2}
+\frac{M \sin^2 \theta}{r} \left ( 82 \cos^2 \theta - 17\right )
+ 81 \cos^4 \theta - 18 \cos^2 \theta - 17
\right \}
\right ] \nonumber \\
&& ~~~~-\frac{\kappa^4}{4W^4} \left [ \frac{3528 M^4 \sin^2
\theta}{r^{16}} \left ( 1 - \frac{2M \sin^2 \theta}{r}\right )
\right ] \ne 0,
\end{eqnarray}
\end{itemize}
which shows clearly that the Kerr metric is not a solution to the
modified HL gravity. Even at the zeroth order of $a$, the equation
(\ref{zerotheq}) is not satisfied, which means that the
Schwarzschild metric is not the solution to the modified HL gravity
even though it is the solution to the Einstein gravity. When all
higher-order terms are turned off ($\kappa \to 0,~\omega \to
\infty$), we find either the Kerr solution or Schwarzschild solution
as in the Einstein gravity.

Lastly, we  show that Eq. (\ref{eom3}) is not satisfied by Kerr
metric in the order-by-order of $a$. The explicit form of
$E_{ij}^{(k)}$ is shown in Appendix \ref{apendixD}.  Eq.
 (\ref{eom3}) can be split into
\begin{itemize}
\item $a^0$-order, $(rr)$ component:
\begin{equation} \label{zerotheq3rr}
\frac{3 \kappa}{2^{5/2}} \left [\frac{M^2}{r^6\sqrt{1-\frac{2M}{r}}} \right ]\ne 0,
\end{equation}
\item $a^0$-order, $(\theta\theta)$ component:
\begin{equation} \label{zerotheq3thth}
-\frac{3 \kappa}{2^{3/2}} \left [\frac{M^2}{r^4} \sqrt{1-\frac{2M}{r}} \right ]\ne 0,
\end{equation}
\item $a^0$-order, $(\phi\phi)$ component:
\begin{equation} \label{zerotheq3phph}
-\frac{3 \kappa}{2^{3/2}} \left [\frac{M^2\sin^2 \theta}{r^4} \sqrt{1-\frac{2M}{r}} \right ]\ne 0,
\end{equation}
\item $a^2$-order, $(rr)$ component:
\begin{eqnarray} \label{2ndeq3rr}
&&
-\frac{3 \kappa}{2^{5/2}} \left [
\frac{M^2}{r^8\left ( 1-\frac{2M}{r}\right )^{3/2}}
\left \{
12 -28 \cos^2\theta 
-\frac{M}{r}\left ( 13 - 46 \cos^2\theta\right )
-\frac{M^2}{r^2} \left ( 20\sin^2\theta \right )
\right \}
\right ]  \nonumber \\
&&
+ \frac{357\kappa^2}{W^4} \left [ \frac{M^3 }{r^{11} \sqrt{1-\frac{2M}{r}}}
\left ( 1-3\cos^2\theta\right ) \right ] 
\ne 0,
\end{eqnarray}
\item $a^2$-order, $(r\theta)$ component:
\begin{eqnarray} \label{2ndeq3rth}
&&
\frac{3\kappa}{2^{5/2}} \left [
\frac{M^2}{r^8 \sqrt { 1-\frac{2M}{r}}}
\left ( 61 - 136 \frac{M}{r} \right )
\right ] \nonumber \\
&& 
-\frac{21\kappa^2}{2W^4} \left [ \frac{M^3 }{r^{10} \sqrt{1-\frac{2M}{r}}}
\left ( 197 -403\cos^2\theta\right ) \right ] 
\ne 0, \nonumber \\
&&
\end{eqnarray}
\item $a^2$-order, $(r\phi)$ component:
\begin{eqnarray} \label{2ndeq3rph}
&&
\frac{9\kappa^{3/2}}{2^{11/4}W^2} \left [
\frac{M^2\cos\theta\sin^2\theta}{r^9}
\left ( 228 - 533 \frac{M}{r} \right )
\right ] 
\ne 0, 
\end{eqnarray}
\item $a^2$-order, $(\theta\theta)$ component:
\begin{eqnarray} \label{2ndeq3thth}
&&
-\frac{3\kappa}{2^{7/2}} \left [
\frac{M^2}{r^6\sqrt { 1-\frac{2M}{r}}}
\left \{
-81 + 73 \cos^2\theta 
+\frac{M}{r}\left ( 354 - 336 \cos^2\theta\right ) 
-\frac{M^2}{r^2} \left ( 380\sin^2\theta \right )
\right \}
\right ]  \nonumber \\
&&
 - \frac{21\kappa^2}{W^4} \left [ \frac{M^3 }{r^{9} \sqrt{1-\frac{2M}{r}}}
\left \{ -15-2\cos^2\theta 
+\frac{M}{r} \left ( 106 - 72 \cos^2\theta\right ) 
-\frac{M^2}{r^2} \left ( 152 \sin^2\theta\right )
\right \} \right ] \nonumber \\
&&
\ne 0,
\end{eqnarray}
\item $a^2$-order, $(\theta\phi)$ component:
\begin{eqnarray} \label{2ndeq3thph}
&&
\frac{3}{4}\frac{3 \kappa^{3/2}}{2^{7/4}W^2} \left [
\frac{M^2\sin^3\theta}{r^9}
\left ( 513 - 2796 \frac{M}{r} + 3625 \frac{M^2}{r^2}\right )
\right ] 
\ne 0, 
\end{eqnarray}
\item $a^2$-order, $(\phi\phi)$ component:
\begin{eqnarray} \label{2ndeq3phph}
&&
\frac{3\kappa}{2^{5/2}} \left [
\frac{M^2 \sin^2\theta}{r^6\sqrt { 1-\frac{2M}{r}}}
\left \{
-81 + 89 \cos^2\theta 
+\frac{M}{r}\left ( 404 - 422 \cos^2\theta\right )
-\frac{M^2}{r^2} \left ( 488 \sin^2\theta \right )
\right \}
\right ]  \nonumber \\
&&
 + \frac{21\kappa^2}{W^4} \left [ \frac{M^3 \sin^2\theta}{r^{9} \sqrt{1-\frac{2M}{r}}}
\left \{ -32 + 49 \cos^2\theta 
+\frac{M}{r} \left ( 140 - 174 \cos^2\theta\right )
-\frac{M^2}{r^2} \left ( 152 \sin^2\theta\right )
\right \} \right ] \nonumber \\
&&
\ne 0, 
\end{eqnarray}
\item $a^4$-order, $(rr)$ component:
\begin{eqnarray} \label{4theq3rr}
&&
\frac{3\kappa}{2^{9/2}} \left [
\frac{M^2}{r^{10}\left ( 1-\frac{2M}{r}\right )^{5/2}}
\left \{
-64 + 1232 \cos^2\theta - 1672 \cos^4\theta \right . \right . \nonumber \\
&&
+\frac{M}{r}\left ( 612 - 6364 \cos^2\theta + 7708\cos^4\theta \right ) 
-\frac{M^2}{r^2}\left ( 1029 - 9406 \cos^2\theta + 10267\cos^4\theta \right ) \nonumber \\
&& \left . \left. 
-\frac{M^3}{r^3}\left ( 868 + 1208 \cos^2\theta - 2076\cos^4\theta \right ) 
+\frac{M^4}{r^4} \left ( 2004 \sin^4\theta \right )
\right \}
\right ]  \nonumber \\
&&
- 3 \frac{\kappa^2}{W^4} \left [ 
\frac{M^3 }{r^{13} \left ( 1-\frac{2M}{r} \right )^{5/2}}
\left \{
679 + 3303 \cos^2\theta - 6898 \cos^4\theta \right . \right . \nonumber \\
&&
-\frac{M}{r}\left ( 5443 + 13482 \cos^2\theta - 30351\cos^4\theta \right ) 
+\frac{M^2}{r^2}\left ( 19826 + 1650 \cos^2\theta - 32664\cos^4\theta \right ) \nonumber \\
&& \left . \left.
-\frac{M^3}{r^3}\left ( 35240 - 48060 \cos^2\theta + 12820\cos^4\theta \right ) 
+\frac{M^4}{r^4} \left ( 23856 \sin^4\theta \right )
\right \}
\right ]  
\ne 0, 
\end{eqnarray}
\item $a^4$-order, $(r\theta)$ component:
\begin{eqnarray} \label{4theq3rth}
&&
\frac{3\kappa}{2^{5/2}} \left [
\frac{M^2\cos\theta\sin\theta}{r^{9}\left ( 1-\frac{2M}{r}\right )^{3/2}}
\left \{
140 - 616 \cos^2\theta  
-\frac{M}{r}\left ( 1002 - 3091 \cos^2\theta  \right ) \right . \right . \nonumber \\
&& \left . \left .
+\frac{M^2}{r^2}\left ( 2598 - 4858 \cos^2\theta \right ) 
-\frac{M^3}{r^3} \left ( 2280 \sin^2\theta \right )
\right \}
\right ]  \nonumber \\
&&
- \frac{3}{2} \frac{\kappa^2}{W^4} \left [ 
\frac{M^3 \cos\theta \sin\theta}{r^{12} \left ( 1-\frac{2M}{r} \right )^{3/2}}
\left \{
-1552 - 18464 \cos^2\theta  
+\frac{M}{r}\left ( 1513 + 80510 \cos^2\theta \right ) \right . \right . \nonumber \\
&& \left .
+\frac{M^2}{r^2}\left ( 17292 - 101207 \cos^2\theta \right ) 
-\frac{M^3}{r^3} \left ( 28086 \sin^2\theta \right )
\right \} \nonumber \\
&& \left .
+\frac{M^4 \sin^2\theta}{r^{14} \left ( 1-\frac{2M}{r} \right )^{3/2}}
\left \{
 1176 
-\frac{M}{r}\left ( 4704 - 2352 \cos^2\theta \right ) 
+\frac{M^2}{r^2} \left ( 4704 \sin^2\theta \right )
\right \} 
\right ] \nonumber \\
&& 
\ne 0, 
\end{eqnarray}
\item $a^4$-order, $(r\phi)$ component:
\begin{eqnarray} \label{4theq3rph}
&&
-\frac{3 \kappa^{3/2}}{2^{11/4}W^2} \left [
\frac{M^2\cos\theta\sin^2\theta}{r^{10}}
\left \{
-1914 + 8010 \cos^2\theta  
+\frac{M}{r}\left ( 12825 - 29888 \cos^2\theta  \right ) \right . \right . \nonumber \\
&& \left . \left .
-\frac{M^2}{r^2} \left ( 17424 \sin^2\theta \right )
\right \}
\right ]  
- \frac{1764\kappa^2}{W^4} \left [ 
\frac{M^4 \sin^2\theta }{r^{14} \sqrt{1-\frac{2M}{r}}} 
\left( 1-\frac{2M}{r}\sin^2\theta\right )
\right ]
\ne 0, 
\end{eqnarray}
\item $a^4$-order, $(\theta\theta)$ component:
\begin{eqnarray} \label{4theq3thth}
&&
\frac{3\kappa}{2^{9/2}} \left [
\frac{M^2}{r^{8}\left ( 1-\frac{2M}{r}\right )^{3/2}}
\left \{
-192 - 2904 \cos^2\theta + 2992 \cos^4\theta \right . \right . \nonumber \\
&&
-\frac{M}{r}\left ( 352 - 22540 \cos^2\theta + 21732\cos^4\theta \right ) 
+\frac{M^2}{r^2}\left ( 10487 - 68870 \cos^2\theta + 57891\cos^4\theta \right ) \nonumber \\
&& \left . \left. 
-\frac{M^3}{r^3}\left ( 31756 + 98280 \cos^2\theta - 66524\cos^4\theta \right ) 
+\frac{M^4}{r^4} \left ( 27468 \sin^4\theta \right )
\right \}
\right ]  \nonumber \\
&&
+ \frac{3\kappa^2}{W^4} \left [ 
\frac{M^3 }{r^{11} \left ( 1-\frac{2M}{r} \right )^{3/2}}
\left \{
501 + 798 \cos^2\theta - 2638 \cos^4\theta \right . \right . \nonumber \\
&&
-\frac{M}{r}\left ( 5443 - 7103 \cos^2\theta - 3805\cos^4\theta \right ) 
+\frac{M^2}{r^2}\left ( 24092 - 54816 \cos^2\theta + 25130\cos^4\theta \right ) \nonumber \\
&& \left .
-\frac{M^3}{r^3}\left ( 47848 - 10962 \cos^2\theta + 61764\cos^4\theta \right ) 
+\frac{M^4}{r^4} \left ( 34776 \sin^4\theta \right )
\right \} \nonumber \\
&&
+ 
\frac{M^4 }{r^{14} \left ( 1-\frac{2M}{r} \right )^{3/2}}
\left \{
-588 \sin^2\theta 
+\frac{M}{r}\left ( 2352 - 3528 \cos^2\theta + 1176\cos^4\theta \right ) \right . \nonumber \\
&& \left . \left .
+\frac{M^2}{r^2} \left ( 2352 \sin^4\theta \right )
\right \} 
\right ]  
\ne 0, 
\end{eqnarray}
\item $a^4$-order, $(\theta\phi)$ component:
\begin{eqnarray} \label{4theq3thph}
&&
\frac{3\kappa}{2^{5/2}} \left [
\frac{M^2\sin^3\theta}{r^{9}}
\left \{
-144 - 10680 \cos^2\theta  
-\frac{M}{r}\left ( 5771 - 72572 \cos^2\theta  \right ) \right . \right . \nonumber \\
&& \left . \left .
+\frac{M^2}{r^2}\left ( 44773 - 140725 \cos^2\theta  \right ) 
-\frac{M^3}{r^3} \left ( 73174 \sin^2\theta \right )
\right \}
\right ] \nonumber \\
&& 
- \frac{1764\kappa^2}{W^4} \left [ 
\frac{M^4 \sin^2\theta }{r^{14} \sqrt{1-\frac{2M}{r}}} 
\left( 1-\frac{2M}{r}\sin^2\theta\right )
\right ]
\ne 0, 
\end{eqnarray}
\item $a^4$-order, $(\phi\phi)$ component:
\begin{eqnarray} \label{4theq3phph}
&&
-\frac{3 \kappa}{2^{9/2}} \left [
\frac{M^2 \sin^2\theta}{r^{8}\left ( 1-\frac{2M}{r}\right )^{3/2}}
\left \{
 48 - 2412 \cos^2\theta + 2468 \cos^4\theta \right . \right . \nonumber \\
&&
-\frac{M}{r}\left ( 1628 - 21852 \cos^2\theta + 20680 \cos^4\theta \right ) \nonumber \\
&&
+\frac{M^2}{r^2}\left ( 12277 - 71578 \cos^2\theta + 57793\cos^4\theta \right ) \nonumber \\
&& \left . \left .
-\frac{M^3}{r^3}\left ( 31812 - 101800 \cos^2\theta + 69988\cos^4\theta \right ) 
+\frac{M^4}{r^4} \left ( 26756 \sin^4\theta \right )
\right \}
\right ]  \nonumber \\
&&
- \frac{3\kappa^2}{W^4} \left [ 
\frac{M^3 \sin^2\theta}{r^{11} \left ( 1-\frac{2M}{r} \right )^{3/2}}
\left \{
165 - 3429 \cos^2\theta + 4603 \cos^4\theta \right . \right . \nonumber \\
&&
-\frac{M}{r}\left ( 1502 - 23791 \cos^2\theta + 27764 \cos^4\theta \right ) \nonumber \\
&&
+\frac{M^2}{r^2}\left ( 6198 - 58678 \cos^2\theta + 58074 \cos^4\theta \right ) \nonumber \\
&& \left . 
-\frac{M^3}{r^3}\left ( 12216 - 58640 \cos^2\theta + 46424 \cos^4\theta \right ) 
+\frac{M^4}{r^4} \left ( 9016 \sin^4\theta \right )
\right \} \nonumber \\
&& \left .
- 
\frac{M^4 \sin^2\theta}{r^{14} \left ( 1-\frac{2M}{r} \right )^{3/2}}
\left \{
588 
-\frac{M}{r}\left ( 2352 + 1178 \cos^2\theta \right ) 
+\frac{M^2}{r^2} \left ( 2352 \sin^4\theta \right )
\right \} 
\right ]  \nonumber \\
&&
\ne 0. 
\end{eqnarray}
\end{itemize}

 As was noted
previously, the higher-order terms violate the Lorentz symmetry.
Hence, it is conjectured that the absence of Kerr solution is
related closely to the violation of the Lorentz-symmetry in the HL
gravity.   In order to explore a connection between Kerr solution
and Lorentz-symmetry, we introduce one known-example. It is well
known that the CS modified gravity could not provide the  Kerr
metric as a fully rotating black hole solution since it cannot
satisfy the Pontryagin constraint.    The action for the CS modified
gravity is given by
\begin{equation}
S_{\rm EH}+S_{\rm CS}+S_{\rm m} = \frac{1}{\kappa^2} \int d^4 x
\sqrt{-g} \left ( R -
 \frac{1}{4}\vartheta~{}^*R R + \kappa^2{\cal L}_m\right ) ,
\label{action-CS}
\end{equation}
where $\vartheta$ is a CS parameter and thus,
$v_{\mu}=\partial_{\mu} \vartheta$ plays a role of the embedding
vector. ${\cal L}_m$ represents the matter whose energy-momentum
tensor is $T^{\mu\nu}$. The equation of motion CS modified gravity
is given by
\begin{equation}
G^{\mu\nu} + C^{\mu\nu} = \kappa^2 T^{\mu\nu},
\label{eom-CS}
\end{equation}
where $C^{\mu\nu}$ is  the Cotton tensor defined as
\begin{equation}
C^{\mu\nu} = \frac{1}{\sqrt{-g}} \frac{\del S_{\rm CS}}{\del g_{\mu\nu}}.
\label{cotton-CS}
\end{equation}
Requiring the Bianchi identity and using the energy-momentum
conservation, we have the condition
\begin{equation}
\nabla^{\mu} C_{\mu\nu} = 0.
\label{pont-eq}
\end{equation}
Considering the relation \begin{equation} \nabla^{\mu} C_{\mu\nu}=
\frac{v_{\nu}}{8}~{}^*RR,\end{equation}
 the Pontryagin constraint
\begin{equation}
{\cal P} ={}^*RR = 0, \label{pont-const}
\end{equation}
should be satisfied for any solution to the CS modified gravity.
However, for the Kerr solution (\ref{kerr-metric}), one has
$\nabla^\mu C_{\mu\nu}=0$, but ${\cal P}$ is not zero as
\begin{equation}
{\cal P}={}^*RR=
\frac{96aM^2r}{\rho^{12}}\cos\theta\Big(r^2-3a^2\cos^2\theta\Big)(3r^2-a^2\cos^2\theta\Big).
\end{equation}
In the limit of $a\to 0$, the Schwarzschild solution is recovered
with ${\cal P}=0$. Therefore, for any finite $a$, the Pontryagin
term is non-vanishing and thus, the Kerr spacetime cannot be a
solution to the CS modified gravity equation.

\section{Discussions}
We have shown that the Kerr metric is not a solution of the modified
HL gravity with $\lambda=1$ and $\Lambda_{\rm W}=0$. This was
performed by checking whether or not three equations are satisfied
for the Kerr metric, according to the $a$-order. It is shown that
the lapse equation (\ref{eom1}) and the metric equation (\ref{eom3})
are not satisfied for the Kerr metric (\ref{kerr-metric}) even at
the zeroth order of rotating parameter $a$. This implies that the
Schwarzschild metric is not a solution to the modified HL gravity because in
the limit of $a \to 0$, the Kerr metric reduces to the Schwarzschild
metric.  The dissatisfaction comes from the presence of higher-order
curvature terms in the modified HL gravity, which may enable us to
carry the power-counting renormalizability out.

 The only
allowable rotating solution to the modified HL gravity is the slowly
rotating KS solution which includes the effect ($\omega$) of
higher-order curvature terms.  We mention again that the slowly
rotating Kerr metric is not the solution to the modified HL gravity
but the Einstein gravity.

In conclusion, the absence of a fully rotating solution in the
modified HL gravity provides another dark-side, in addition to the
strong coupling issue because  astrophysical black holes may be
considered as the Kerr black hole~\cite{Bambi}.

\section*{Acknowledgement}
This work was supported by  the National Research Foundation of Korea
(NRF) grant funded by the Korea government (MEST) (No.2011-0027293).

\appendix
\section{Ricci tensor and scalar}
We  show the explicit form of  $R_{ij}$ and $R$ for the Kerr metric.

 \label{apendixA} Ricci tensor is given by
\begin{equation} R_{ij}=\begin{pmatrix}
         R_{rr} & R_{r\theta} & 0 \\
         R_{r\theta} & R_{\theta\theta} & 0 \\
         0 & 0 & R_{\phi\phi}
        \end{pmatrix},
\label{R_ij_kerr1}
\end{equation}
where
\begin{eqnarray}
R_{rr} &=& -M\frac{Q_{rr}}{P_{rr}}, \\
Q_{rr} &=&
2\,{r}^{11} -16\, {r}^{7}{a}^{4}\cos^2 \theta
-22\, {r}^{3}{a}^{8}\cos^4 \theta
+3\, r{a}^{10}\cos^6 \theta
+3\, {r}^{5}{a}^{6}\cos^6 \theta  \nonumber \\
&&
+8\, {r}^{7}{a}^{4}+7\,{r}^{9}{a}^{2}+3\,{r}^{5}{a}^{6}
+6\, {r}^{3}{a}^{8}\cos^6 \theta
+7\, M{r}^{8}{a}^{2}\cos^2 \theta
+30\, M{r}^{2}{a}^{8} \cos^4 \theta  \nonumber \\
&&
+16\,M{r}^{6}{a}^{4}\cos^4 \theta
-14\,M{r}^{2}{a}^{8}\cos^6 \theta
-16\,M{r}^{2}{a}^{8}\cos^2 \theta
+12\,{M}^{2}{r}^{3}{a}^{6}\cos^6 \theta  \nonumber \\
&&
-11\, M{r}^{4}{a}^{6}\cos^6 \theta
-24\, {M}^{2}{r}^{3}{a}^{6}\cos^4 \theta
+35\, M{r}^{4}{a}^{6}\cos^4 \theta
-4\, {M}^{2}{r}^{5}{a}^{4}\cos^4 \theta  \nonumber \\
&&
-29\, M{r}^{4}{a}^{6} \cos^2 \theta
+12\, {M}^{2}{r}^{3}{a}^{6} \cos^2 \theta
-18\, M{r}^{6}{a}^{4} \cos^2 \theta
+8\, {M}^{2}{r}^{5}{a}^{4}\cos^2 \theta  \nonumber \\
&&
-7\, M{r}^{8}{a}^{2}+2\,M{r}^{6}{a}^{4}-4\,{M}^{2}{r}^{5}{a}^{4}
+5\, M{r}^{4}{a}^{6}
-4\, {r}^{7}{a}^{4} \cos^4 \theta
-17\, {r}^{5}{a}^{6}\cos^2 \theta  \nonumber \\
&&
-5\, {r}^{9}{a}^{2}\cos^2 \theta
-17\, {r}^{5}{a}^{6}\cos^4 \theta
-9\, r{a}^{10}\cos^4 \theta
-6\,{r}^{3}{a}^{8}\cos^2 \theta  \nonumber \\
&&
+M{a}^{10}\cos^6 \theta
-M{a}^{10}\cos^4 \theta  ,
\end{eqnarray}
\begin{eqnarray}
P_{rr} &=&
-{r}^{14}\left(1-\frac{2\,M}{r}\right)-{r}^{8}{a}^{6}
+8\,{M}^{2}{r}^{10}{a}^{2} +8\,{M}^{3}{r}^{7}{a}^{4}
-3\, {r}^{2}{a}^{12} \cos^8 \theta  \nonumber \\
&&
- {r}^{6}{a}^{8}\cos^8 \theta
-4\, {r}^{2}{a}^{12}\cos^6 \theta
-12\,{r}^{10}{a}^{4} \cos^2 \theta  \nonumber \\
&&
-12\,{r}^{8}{a}^{6} \cos^2 \theta
-18\,{r}^{8}{a}^{6} \cos^4 \theta
-18\,{r}^{6}{a}^{8} \cos^4 \theta
-12\,{r}^{6}{a}^{8} \cos^6 \theta \nonumber \\
&&
-12\,{a}^{10} \left( \cos \left( \theta \right)  \right)
^{6}{r}^{4} -4\, \left( \cos \left( \theta \right)  \right)
^{2}{a}^{8}{r}^{6} -6\, \left( \cos \left( \theta \right)  \right)
^{4}{a}^{10}{r}^{4}
-4\,{a}^{2} \left( \cos \left( \theta \right)  \right) ^{2}{r}^{12} \nonumber \\
&&
-6\, {r}^{10}{a}^{4} \cos^4 \theta
-4\, {r}^{8}{a}^{6} \cos^6 \theta
-3\, {r}^{4}{a}^{10}\cos^{8} \theta
-12\,M{r}^{7}{a}^{6} \cos^2 \theta  \nonumber \\
&&
+12\, M{r}^{3}{a}^{10} \cos^6 \theta
-12\, M{r}^{5}{a}^{8} \cos^2 \theta
-12\, M{r}^{3}{a}^{10}\cos^4 \theta
+36\, M{r}^{5}{a}^{8} \cos^6 \theta  \nonumber \\
&&
+24\, {M}^{2}{r}^{8}{a}^{4} \cos^2 \theta
+12\,M{r}^{11}{a}^{2} \cos^2 \theta
+36\, {M}^{2}{r}^{4}{a}^{8}\cos^4 \theta
+12\, M{r}^{9}{a}^{4} \cos^2 \theta  \nonumber \\
&&
+36\,M{r}^{7}{a}^{6} \cos^4 \theta
-16\, {M}^{3}{r}^{3}{a}^{8} \cos^6 \theta
+8\,{M}^{3}{r}^{3}{a}^{8}\cos^8 \theta
+16\, {M}^{3}{r}^{5}{a}^{6} \cos^6 \theta  \nonumber \\
&&
-16\,{M}^{3}{r}^{7}{a}^{4} \cos^2 \theta
+24\,{M}^{2}{r}^{6}{a}^{6} \cos^2 \theta
-8\,{M}^{2}{r}^{10}{a}^{2} \cos^2 \theta
+20\, M{r}^{7}{a}^{6} \cos^6 \theta  \nonumber \\
&&
-16\, {M}^{2}{r}^{4}{a}^{8} \cos^6 \theta
-8\,{M}^{2}{r}^{4}{a}^{8} \cos^2 \theta
-4\,{M}^{2}{r}^{2}{a}^{10} \cos^4  \theta
+24\, M{r}^{9}{a}^{4} \cos^4 \theta \nonumber \\
&&
+12\, {M}^{2}{r}^{6}{a}^{6} \cos^4 \theta
+16\, {M}^{2}{r}^{2}{a}^{10} \cos^6 \theta
-32\,{M}^{2}{r}^{6}{a}^{6} \cos^6 \theta
+8\, {M}^{3}{r}^{3}{a}^{8} \cos^4 \theta \nonumber \\
&&
+16\, {M}^{3}{r}^{5}{a}^{6} \cos^2 \theta
-28\, {M}^{2}{r}^{8}{a}^{4} \cos^4 \theta
-32\, {M}^{3}{r}^{5}{a}^{6} \cos^4 \theta
+8\, {M}^{3}{r}^{7}{a}^{4} \cos^4 \theta \nonumber \\
&&
-4\,Mr{a}^{12} \cos^6 \theta
+6\, Mr{a}^{12} \cos^8 \theta
+12\,M{r}^{3}{a}^{10} \cos^8  \theta
-12\, {M}^{2}{r}^{2}{a}^{10} \cos^8 \theta \nonumber \\
&&
+6\, M{r}^{5}{a}^{8} \cos^8 \theta
-12\, {M}^{2}{r}^{4}{a}^{8} \cos^8\theta
-6\,M{r}^{9}{a}^{4}+4\,{M}^{2}{r}^{8}{a}^{4}
-{a}^{14} \cos^8 \theta  \nonumber \\
&&
-4\,{M}^{2}{r}^{6}{a}^{6}-4\,M{r}^{7}{a}^{6} -3\,{r}^{10}{a}^{4}
-3\,{r}^{12}{a}^{2},
\end{eqnarray}
\begin{eqnarray}
R_{r\theta} &=& -M{a}^{2} \sin \theta \cos \theta \frac{Q_{r\theta}}{P_{r\theta}}, \\
Q_{r\theta} &=&
9\,{r}^{8}+4\,Mr{a}^{6} \cos^4 \theta
+8\, M{r}^{3}{a}^{4} \cos^4 \theta
-3\, {r}^{4}{a}^{4} \cos^4 \theta
-6\,{r}^{2}{a}^{6} \cos^4  \theta  \nonumber \\
&&
-3\, {a}^{8} \cos^4 \theta
-12\,M{r}^{5}{a}^{2} \cos^2 \theta
+6\,{r}^{6}{a}^{2} \cos^2 \theta
-4\,Mr{a}^{6} \cos^2  \theta \nonumber \\
&&
+6\,{r}^{2}{a}^{6} \cos^2 \theta
+12\,{r}^{4}{a}^{4} \cos^2 \theta
-24\,M{r}^{3}{a}^{4} \cos^2  \theta
+18\,{r}^{6}{a}^{2} \nonumber \\
&&
+16\,{a}^{4}M{r}^{3} +12\,{a}^{2}M{r}^{5}
+9\,{a}^{4}{r}^{4} ,
\end{eqnarray}
\begin{eqnarray}
P_{r\theta} &=&
r^{12}+ 4\,{M}^{2}{r}^{2}{a}^{8} \cos^4 \theta
+8\, {M}^{2}{r}^{4}{a}^{6} \cos^2 \theta
+{r}^{8}{a}^{4}+2\,{r}^{10}{a}^{2}+4\,M{r}^{7}{a}^{4}
+4\,{M}^{2}{r}^{6}{a}^{4} \nonumber \\
&&
+4\,M{r}^{9}{a}^{2}
-4\, M{r}^{3}{a}^{8} \cos^8 \theta
-4\, Mr{a}^{10} \cos^8 \theta
+4\, {M}^{2}{r}^{2}{a}^{8} \cos^8 \theta  \nonumber \\
&&
-8\, M{r}^{3}{a}^{8} \cos^6  \theta
+4\, Mr{a}^{10} \cos^6 \theta
-8\, {M}^{2}{r}^{2}{a}^{8} \cos^6 \theta
-12\, M{r}^{5}{a}^{6} \cos^6 \theta \nonumber \\
&&
+8\, {M}^{2}{r}^{4}{a}^{6} \cos^6 \theta
+12\, M{r}^{3}{a}^{8} \cos^4 \theta
-16\, {M}^{2}{r}^{4}{a}^{6} \cos^4 \theta
-12\, M{r}^{7}{a}^{4} \cos^4 \theta \nonumber \\
&&
+4\, {M}^{2}{r}^{6}{a}^{4} \cos^4  \theta
+8\, M{r}^{7}{a}^{4} \cos^2 \theta
-8\, {M}^{2}{r}^{6}{a}^{4} \cos^2 \theta
+12\, M{r}^{5}{a}^{6} \cos^2 \theta \nonumber \\
&&
-4\, M{r}^{9}{a}^{2}  \cos^2 \theta
+ {r}^{4}{a}^{8} \cos^8  \theta
+2\, {r}^{2}{a}^{10} \cos^8  \theta  \nonumber \\
&&
+8\, {r}^{4}{a}^{8} \cos^6 \theta
+4\, {r}^{2}{a}^{10} \cos^6 \theta
+4\, {r}^{6}{a}^{6} \cos^6 \theta
+12\, {r}^{6}{a}^{6} \cos^4 \theta  \nonumber \\
&&
+6\, {r}^{4}{a}^{8} \cos^4 \theta
+6\, {r}^{8}{a}^{4} \cos^4 \theta
+4\, {r}^{6}{a}^{6} \cos^2 \theta
+8\, {r}^{8}{a}^{4} \cos^2 \theta  \nonumber \\
&&
+4\, {r}^{10}{a}^{2} \cos^2 \theta
+ {a}^{12} \cos^8 \theta ,
\end{eqnarray}
\begin{eqnarray}
R_{\theta\theta} &=&- Mr\frac{Q_{\theta\theta}}{P_{\theta\theta}}, \\
Q_{\theta\theta} &=&
r^{10}-22\,M{r}^{3}{a}^{6} \cos^2 \theta
+34\,Mr{a}^{8} \cos^4 \theta
+16\,{M}^{2}{r}^{2}{a}^{6} \cos^6 \theta
-32\, {M}^{2}{r}^{2}{a}^{6} \cos^4 \theta  \nonumber \\
&&
+16\, {M}^{2}{r}^{4}{a}^{4} \cos^2 \theta
+16\, {M}^{2}{r}^{2}{a}^{6} \cos^2 \theta
-14\, Mr{a}^{8} \cos^2 \theta
-8\, {M}^{2}{r}^{4}{a}^{4} \cos^4 \theta \nonumber \\
&&
-8\,{M}^{2}{r}^{4}{a}^{4} +6\,M{r}^{3}{a}^{6} +5\,{r}^{8}{a}^{2}
-2\,M{r}^{7}{a}^{2}
+4\,M{r}^{5}{a}^{4} \nonumber \\
&&
-4\,{r}^{8}{a}^{2} \cos^2 \theta
+{r}^{6}{a}^{4} \cos^4 \theta
+6\,{r}^{4}{a}^{6} \cos^6 \theta
+7\,{r}^{6}{a}^{4}+3\,{r}^{4}{a}^{6} \nonumber \\
&&
-6\,M{r}^{5}{a}^{4} \cos^2 \theta
+36\,M{r}^{3}{a}^{6} \cos^4 \theta
+2\, M{r}^{5}{a}^{4} \cos^4 \theta
-20\, M{r}^{3}{a}^{6} \cos^6 \theta \nonumber \\
&&
+2\, M{r}^{7}{a}^{2} \cos^2 \theta
-20\, Mr{a}^{8} \cos^6 \theta
+6\,{a}^{10} \cos^6 \theta
-9\, {a}^{10} \cos^4 \theta  \nonumber \\
&&
-14\,{r}^{6}{a}^{4} \cos^2 \theta
-7\,{r}^{4}{a}^{6} \cos^4 \theta
+12\,{r}^{2}{a}^{8} \cos^6 \theta
-6\, {r}^{2}{a}^{8} \cos^2 \theta  \nonumber \\
&&
-16\, {r}^{4}{a}^{6} \cos^2 \theta
-17\, {r}^{2}{a}^{8} \cos^4 \theta ,
\end{eqnarray}
\begin{eqnarray}
P_{\theta\theta} &=&
r^{12}+ 4\, {M}^{2}{r}^{2}{a}^{8} \cos^4 \theta
+8\, {M}^{2}{r}^{4}{a}^{6} \cos^2 \theta
+{r}^{8}{a}^{4}+2\,{r}^{10}{a}^{2}+4\,M{r}^{7}{a}^{4} \nonumber \\
&&
+4\,{M}^{2}{r}^{6}{a}^{4} +4\,M{r}^{9}{a}^{2}
-4\, M{r}^{3}{a}^{8} \cos^8 \theta
-4\, Mr{a}^{10} \cos^8 \theta  \nonumber \\
&&
+4\, {M}^{2}{r}^{2}{a}^{8} \cos^8 \theta
-8\, M{r}^{3}{a}^{8} \cos^6 \theta
+4\, Mr{a}^{10} \cos^6 \theta
-8\, {M}^{2}{r}^{2}{a}^{8} \cos^6 \theta  \nonumber \\
&&
-12\, M{r}^{5}{a}^{6}\cos^6 \theta
+8\, {M}^{2}{r}^{4}{a}^{6} \cos^6 \theta
+12\, {M}^{2}{r}^{3}{a}^{8} \cos^4 \theta
-16\, {M}^{2}{r}^{4}{a}^{6} \cos^4 \theta  \nonumber \\
&&
-12\, M{r}^{7}{a}^{4} \cos^4 \theta
+4\, {M}^{2}{r}^{6}{a}^{4} \cos^4 \theta
+8\, M{r}^{7}{a}^{4} \cos^2 \theta
-8\, {M}^{2}{r}^{6}{a}^{4} \cos^2 \theta \nonumber \\
&&
+12\, M{r}^{5}{a}^{6} \cos^2 \theta
-4\, M{r}^{9}{a}^{2} \cos^2 \theta
+ {r}^{4}{a}^{8} \cos^8 \theta
+2\, {r}^{2}{a}^{10} \cos^8 \theta  \nonumber \\
&&
+8\,{r}^{4}{a}^{8} \cos^6 \theta
+4\, {r}^{2}{a}^{10} \cos^6 \theta
+4\, {r}^{6}{a}^{6} \cos^6 \theta
+12\, {r}^{6}{a}^{6} \cos^4  \theta \nonumber \\
&&
+6\, {r}^{4}{a}^{8} \cos^4 \theta
+6\, {r}^{8}{a}^{4} \cos^4 \theta
+4\, {r}^{6}{a}^{6} \cos^2 \theta
+8\, {r}^{8}{a}^{4} \cos^2 \theta  \nonumber \\
&&
+4\, {r}^{10}{a}^{2} \cos^2 \theta
+ {a}^{12} \cos^8 \theta ,
\end{eqnarray}
\begin{eqnarray}
R_{\phi\phi} &=&- M \sin^2 \theta \frac{Q_{\phi\phi}}{P_{\phi\phi}}, \\
Q_{\phi\phi} &=&
-{r}^{11}+2\, {r}^{7}{a}^{4} \cos^2 \theta
+5\, {r}^{3}{a}^{8}\cos^4 \theta
+3\, r{a}^{10} \cos^6 \theta
+3\, {r}^{5}{a}^{6} \cos^6  \theta  \nonumber \\
&&
-{r}^{7}{a}^{4} -2\,{r}^{9}{a}^{2}
+6\, {r}^{3}{a}^{8} \cos^6  \theta
+13\, M{r}^{8}{a}^{2} \cos^2 \theta \nonumber \\
&&
+20\,M{r}^{2}{a}^{8} \cos^4 \theta
-2\, M{r}^{6}{a}^{4} \cos^4 \theta
-18\, M{r}^{2}{a}^{8} \cos^6 \theta
-2\, M{r}^{2}{a}^{8} \cos^2 \theta  \nonumber \\
&&
+20\, {M}^{2}{r}^{3}{a}^{6} \cos^6  \theta
-15\, M{r}^{4}{a}^{6} \cos^6 \theta
-40\, {M}^{2}{r}^{3}{a}^{6} \cos^4  \theta
+7\, M{r}^{4}{a}^{6} \cos^4  \theta  \nonumber \\
&&
-4\, {M}^{2}{r}^{5}{a}^{4} \cos^4 \theta
+9\, M{r}^{4}{a}^{6} \cos^2 \theta
+20\, {M}^{2}{r}^{3}{a}^{6} \cos^2 \theta
+12\, M{r}^{6}{a}^{4} \cos^2 \theta  \nonumber \\
&&
+8\, {M}^{2}{r}^{5}{a}^{4} \cos^2  \theta
-13\,M{r}^{8}{a}^{2} -10\,M{r}^{6}{a}^{4}-4\,{M}^{2}{r}^{5}{a}^{4}
-M{r}^{4}{a}^{6} \nonumber \\
&&
+5\, {r}^{7}{a}^{4} \cos^4  \theta
+ {r}^{5}{a}^{6} \cos^2  \theta
+{r}^{9}{a}^{2} \cos^2  \theta
+10\, {r}^{5}{a}^{6} \cos^4  \theta  \nonumber \\
&&
+ M{a}^{10} \cos^6 \theta
- M{a}^{10} \cos^4 \theta  ,
\end{eqnarray}
\begin{eqnarray}
P_{\phi\phi} &=&
-r^{12}+ 2\, Mr{a}^{10} \cos^{10} \theta
-{r}^{10}{a}^{2} -2\,M{r}^{9}{a}^{2}
+8\, M{r}^{3}{a}^{8} \cos^8 \theta  \nonumber \\
&&
-2\, Mr{a}^{10} \cos^8  \theta
-8\, M{r}^{3}{a}^{8} \cos^6  \theta
+12\,M{r}^{5}{a}^{6} \cos^6  \theta
+8\, M{r}^{7}{a}^{4} \cos^4 \theta  \nonumber \\
&&
-8\, M{r}^{7}{a}^{4} \cos^2 \theta
+2\, M{r}^{9}{a}^{2} \cos^2 \theta
-5\, {r}^{4}{a}^{8} \cos^8  \theta
-5\, {r}^{2}{a}^{10} \cos^8  \theta \nonumber \\
&&
-10\, {r}^{4}{a}^{8}\cos^6  \theta
-10\, {r}^{6}{a}^{6} \cos^6  \theta
-10\, {r}^{6}{a}^{6} \cos^4  \theta
-10\, {r}^{8}{a}^{4} \cos^4  \theta \nonumber \\
&&
-5\, {r}^{8}{a}^{4} \cos^2  \theta
-5\, {r}^{10}{a}^{2} \cos^2  \theta
-12\, M{r}^{5}{a}^{6} \cos^4  \theta  \nonumber \\
&&
- {a}^{12} \cos^{10} \theta
- {r}^{2}{a}^{10} \cos^{10} \theta .
\end{eqnarray}

Ricci scalar is given by
\begin{eqnarray}
R &=& -2 M^2 a^2 \sin^2 \theta  \frac{Q_R}{P_R}, \\
Q_R &=& 9\,{r}^{8}-3\, {r}^{4}{a}^{4}\cos^4 \theta + {a}^{8} \cos^4
\theta -6\,{r}^{2}{a}^{6} \cos^4 \theta
+8\, M{r}^{3}{a}^{4} \cos^4 \theta  \nonumber \\
&& -8\,M{r}^{3}{a}^{4} \cos^2 \theta +2\,{r}^{2}{a}^{6} \cos^2
\theta +6\,{r}^{6}{a}^{2} \cos^2 \theta +{a}^{4}{r}^{4}
+6\,{r}^{6}{a}^{2} ,
\end{eqnarray}
\begin{eqnarray}
P_R &=&r^{14}+ 24\,M{r}^{5}{a}^{8} \cos^4 \theta -4\,
M{r}^{3}{a}^{10} \cos^{10} \theta -4\, Mr{a}^{12}\cos^{10} \theta
+4\, {M}^{2}{r}^{2}{a}^{10}\cos^{10} \theta  \nonumber \\
&& +5\, {r}^{2}{a}^{12}\cos^8 \theta +5\, {r}^{6}{a}^{8}\cos^8
\theta + {a}^{14}\cos^{10} \theta \nonumber \\
&& +2\, {r}^{2}{a}^{12} \cos^{10} \theta + {r}^{4}{a}^{10}\cos^{10}
\theta +4\,M{r}^{11}{a}^{2} +10\,{r}^{10}{a}^{4} \cos^2 \theta
+5\,{r}^{8}{a}^{6} \cos^2 \theta  \nonumber \\
&& +20\,{r}^{8}{a}^{6} \cos^4 \theta +10\,{r}^{6}{a}^{8} \cos^4
\theta +20\,{r}^{6}{a}^{8} \cos^6 \theta
+10\,{r}^{4}{a}^{10} \cos^6 \theta  \nonumber \\
&& +5\,{r}^{12}{a}^{2} \cos^2 \theta +10\,{r}^{10}{a}^{4} \cos^4
\theta +10\,{a}^{6} \cos^6 \theta {r}^{8}
+10\, {r}^{4}{a}^{10}\cos^{8} \theta  \nonumber \\
&& +16\,M{r}^{7}{a}^{6} \cos^2 \theta +16\,M{r}^{3}{a}^{10}  \cos^6
\theta -8\,M{r}^{5}{a}^{8}  \cos^6 \theta
-8\,{M}^{2}{r}^{8}{a}^{4}  \cos^2 \theta  \nonumber \\
&& -4\,M{r}^{11}{a}^{2} \cos^2 \theta +12\, {M}^{2}{r}^{4}{a}^{8}
\cos^4 \theta +12\,M{r}^{9}{a}^{4} \cos^2 \theta
+8\,M{r}^{7}{a}^{6} \cos^4 \theta  \nonumber \\
&& +12\,{a}^{6}{M}^{2}{r}^{6} \cos^2 \theta -24\,
M{r}^{7}{a}^{6}\cos^6 \theta -24\, {M}^{2}{r}^{4}{a}^{8}\cos^6
\theta
-16\, M{r}^{9}{a}^{4}\cos^4 \theta  \nonumber \\
&& -24\, {M}^{2}{r}^{6}{a}^{6}\cos^4 \theta +4\,
{M}^{2}{r}^{2}{a}^{10}\cos^6 \theta +12\,{M}^{2}{r}^{6}{a}^{6}
\cos^6 \theta
+4\, {M}^{2}{r}^{8}{a}^{4}\cos^4 \theta  \nonumber \\
&& +4\, Mr{a}^{12} \cos^8 \theta -12\, M{r}^{3}{a}^{10} \cos^8
\theta -8\, {M}^{2}{r}^{2}{a}^{10} \cos^8 \theta
-16\, M{r}^{5}{a}^{8} \cos^8 \theta  \nonumber \\
&& +12\, {M}^{2}{r}^{4}{a}^{8} \cos^8 \theta +4\,M{r}^{9}{a}^{4}
+4\,{M}^{2}{r}^{8}{a}^{4} +{r}^{10}{a}^{4} +2\,{r}^{12}{a}^{2}.
\end{eqnarray}
\section{ Cotton tensor}
\label{apendixB}We  show the explicit form of $C_{r\phi}$ and
$C_{\theta\phi}$ for the Kerr metric.  $F_{r\phi}$, $N_{r\theta}$,
$D_{r\theta}$, $N_{\theta\phi}$, and $D_{23}$ appeared  in Eqs.
(\ref{c12}) and (\ref{c23}) are given by
\begin{eqnarray}
F_{r\phi}^2&=& \frac {{r}^{2}+{a}^{2}-2\,Mr}{ \left( {r}^{2}+{a}^{2}
\cos^2 \theta  \right) \left( {r}^{4}+{r}^{2}{a}^{2}+2\,Mr{a}^{2}
\sin^2 \theta
 +{r}^{2}{a}^{2} \cos^2 \theta + {a}^{4}\cos^2 \theta
  \right) }
, \\
N_{r\phi} &=&-21\,{r}^{14}
 -30\,M{r}^{11}{a}^{2}-128\,M{r}^{9}{a}^{4}-94\,M{r}^{7}{a}^{6}
-4\,M{r}^{5}{a}^{8}-12\,{M}^{2}{r}^{8}{a}^{4}
 -84\,{r}^{12}{a}^{2} \nonumber \\
&&
  -3\,{r}^{6}{a}^{8}-4\, {r}^{6}{a}^{8} \cos^8 \theta -3\,{a}^{14} \cos^6 \theta
+2\,{a}^{14} \cos^8 \theta
 -52\,{r}^{6}{M}^{2}{a}^{6}
-104\,{r}^{10}{a}^{4} \nonumber \\
&&
-44\,{r}^{8}{a}^{6}-10\,
 {r}^{4}{a}^{10}\cos^8 \theta -4\,
 {a}^{12}{r}^{2}\cos^8 \theta +102\,
 {r}^{8}{a}^{6}\cos^4 \theta
+60\,{r}^{6}{a}^{8}\cos^4 \theta \nonumber \\
&&
+2\, {r}^{4}{a}^{10} \cos^4 \theta +53\,
 {r}^{10}{a}^{4} \cos^4 \theta -9\,
 {r}^{2}{a}^{12}\cos^4 \theta
  -70\,{r}^{10}{a}^{4} \cos^2 \theta
-124\,{r}^{8}{a}^{6} \cos^2 \theta \nonumber \\
&&
-66\,{r}^{6}{a}^{8} \cos^2 \theta
 -3\,{r}^{12}{a}^{2}\cos^2 \theta
 -9\,{r}^{4}{a}^{10}\cos^2 \theta
 +78\,{r}^{6}{a}^{8}\cos^6 \theta
 +76\,{r}^{4}{a}^{10}\cos^6 \theta \nonumber \\
&&
+26\,{r}^{2}{a}^{12}\cos^6 \theta
 +31\,{r}^{8}{a}^{6}\cos^6 \theta
 -98\,M{r}^{9}{a}^{4}\cos^4 \theta
 +34\,M{r}^{7}{a}^{6}\cos^4 \theta \nonumber \\
&&
+190\,M{r}^{5}{a}^{8}\cos^4 \theta
 +30\,M {r}^{11}{a}^{2} \cos^2 \theta
 +226\,M {r}^{9}{a}^{4} \cos^2 \theta
 +170\,M {r}^{7}{a}^{6} \cos^2 \theta \nonumber \\
&&
-110\,M {r}^{7}{a}^{6} \cos^6 \theta
 -202\,M {r}^{5}{a}^{8} \cos^6 \theta
 -114\,M{r}^{3}{a}^{10} \cos^6 \theta
 +102\,M{r}^{3}{a}^{10} \cos^4 \theta \nonumber \\
&&
 -2\,M{r}^{5}{a}^{8} \cos^2 \theta
 +10\,Mr{a}^{12} \cos^6 \theta
 -12\,{M}^{2}{r}^{8}{a}^{4} \cos^4 \theta
 -244\,{M}^{2}{r}^{6}{a}^{6} \cos^4 \theta \nonumber \\
&&
 -180\,{M}^{2}{r}^{4}{a}^{8} \cos^4 \theta
 +24\,{M}^{2}{r}^{8}{a}^{4} \cos^2 \theta
 +200\,{M}^{2}{r}^{6}{a}^{6} \cos^2 \theta
 +96\,{M}^{2}{r}^{6}{a}^{6} \cos^6 \theta \nonumber \\
&&
 +120\,{M}^{2}{r}^{4}{a}^{8} \cos^6 \theta
 -8\,{M}^{2}{r}^{2}{a}^{10} \cos^6 \theta
 +18\,M{r}^{5}{a}^{8} \cos^8 \theta
 +20\,M{r}^{3}{a}^{10} \cos^8 \theta \nonumber \\
&&
 -6\,Mr{a}^{12} \cos^8 \theta
 -20\,{M}^{2}{r}^{4}{a}^{8} \cos^8 \theta
 +4\,{M}^{2}{r}^{2}{a}^{10} \cos^8 \theta
 +4\, {M}^{2}{r}^{2}{a}^{10} \cos^4 \theta \nonumber \\
&&
 -4\, Mr{a}^{12}\cos^4 \theta
 -8\, M{r}^{3}{a}^{10} \cos^2 \theta
 +80\,{M}^{2}{r}^{4}{a}^{8} \cos^2 \theta,
\end{eqnarray}
\begin{eqnarray}
D_{r\phi}&=&-r^{20}
-3\,{r}^{18}{a}^{2}-3\,{r}^{16}{a}^{4}-{r}^{14}{a}^{6}
 -{a}^{20}\cos^{14} \theta
 -7\,{r}^{18}{a}^{2} \cos^2 \theta
 -105\,{r}^{8}{a}^{12} \cos^8 \theta  \nonumber \\
&&
 -35\,{r}^{14}{a}^{6} \cos^6 \theta
 -21\,{r}^{16}{a}^{4} \cos^4
 -63\,{r}^{14}{a}^{6} \cos^4 \theta
 -21\,{r}^{10}{a}^{10} \cos^4 \theta
 -105\,{r}^{12}{a}^{8} \cos^6 \theta \nonumber \\
&&
 -35\,{r}^{8}{a}^{12} \cos^6 \theta
 -105\,{r}^{10}{a}^{10} \cos^8 \theta
 -35\,{r}^{6}{a}^{14} \cos^8 \theta
 -35\,{r}^{12}{a}^{8} \cos^8 \theta \nonumber \\
&&
 -63\,{r}^{6}{a}^{14} \cos^{10} \theta
 -21\,{r}^{10}{a}^{10} \cos^{10} \theta
 -21\,{r}^{14}{a}^{6} \cos^2 \theta
 -63\,{r}^{12}{a}^{8} \cos^4 \theta \nonumber \\
&&
 -105\,{r}^{10}{a}^{10} \cos^6 \theta
 -63\,{r}^{8}{a}^{12} \cos^{10} \theta
 -21\,{r}^{4}{a}^{16} \cos^{10} \theta
 -21\,{r}^{16}{a}^{4} \cos^2 \theta \nonumber \\
&&
 -7\,{r}^{12}{a}^{8} \cos^2 \theta
 -6\,M{r}^{17}{a}^{2}
 -12\,M{r}^{15}{a}^{4}
 -6\,M{r}^{13}{a}^{6}
 -12\,{M}^{2}{r}^{14}{a}^{4}
 -12\,{M}^{2}{r}^{12}{a}^{6} \nonumber \\
&&
 -8\,{M}^{3}{r}^{11}{a}^{6}
 - {r}^{6}{a}^{14} \cos^{14} \theta
 -3\, {r}^{2}{a}^{18}\cos^{14} \theta
 -3\, {r}^{4}{a}^{16}\cos^{14} \theta
 -7\, {r}^{8}{a}^{12}\cos^{12} \theta \nonumber \\
&&
 -21\, {r}^{6}{a}^{14}\cos^{12} \theta
 -21\, {r}^{4}{a}^{16}\cos^{12} \theta
 -7\, {r}^{2}{a}^{18}\cos^{12} \theta
 +6\, M{r}^{5}{a}^{14}\cos^{14} \theta \nonumber \\
&&
 +8\, {M}^{3}{r}^{3}{a}^{14}\cos^{14} \theta
 +12\,M{r}^{3}{a}^{16} \cos^{14} \theta
 -12\,{M}^{2}{r}^{2}{a}^{16} \cos^{14} \theta
 +6\, Mr{a}^{18}\cos^{18} \theta \nonumber \\
&&
 -12\, {M}^{2}{r}^{4}{a}^{14}\cos^{14} \theta
 +36\, M{r}^{7}{a}^{12}\cos^{12} \theta
 +66\, M{r}^{5}{a}^{14}\cos^{12} \theta
 -60\, {M}^{2}{r}^{6}{a}^{12}\cos^{12} \theta \nonumber \\
&&
 +24\, M{r}^{3}{a}^{16}\cos^{12} \theta
 -36\, {r}^{4}{a}^{14}{M}^{2}\cos^{12} \theta
 +32\, {r}^{5}{a}^{12}{M}^{3}\cos^{12} \theta
 -6\, Mr{a}^{18} \cos^{12} \theta \nonumber \\
&&
 +24\, {M}^{2}{r}^{2}{a}^{16} \cos^{12} \theta
 -24\, {M}^{3}{r}^{3}{a}^{14}\cos^{12} \theta
 +90\, M{r}^{9}{a}^{10}\cos^{10} \theta
 +144\, M{r}^{7}{a}^{12}\cos^{10} \theta \nonumber \\
&&
 -120\, {M}^{2}{r}^{8}{a}^{10} \cos^{10} \theta
 +18\, M{r}^{5}{a}^{14}\cos^{10} \theta
 +48\, {M}^{3}{r}^{7}{a}^{10}\cos^{10} \theta
 -36\,M{r}^{3}{a}^{16} \cos^{10} \theta \nonumber \\
&&
 +108\, {M}^{2}{r}^{4}{a}^{14}\cos^{10} \theta
 -96\, {M}^{3}{r}^{5}{a}^{12}\cos^{10} \theta
 -12\, {M}^{2}{r}^{2}{a}^{16}\cos^{10} \theta
 +24\, {M}^{3}{r}^{3}{a}^{14}\cos^{10} \theta \nonumber \\
&&
 +120\, M{r}^{11}{a}^{8}\cos^8 \theta
 +150\, M{r}^{9}{a}^{10} \cos^8 \theta
 -120\, {M}^{2}{r}^{10}{a}^{8}\cos^8 \theta
 -60\, M{r}^{7}{a}^{12}\cos^8 \theta \nonumber \\
&&
 +120\, {M}^{2}{r}^{8}{a}^{10}\cos^8 \theta
 -90\, M{r}^{5}{a}^{14} \cos^8 \theta
 +180\, {M}^{2}{r}^{6}{a}^{12}\cos^8 \theta
 +32\, {M}^{3}{r}^{9}{a}^{8} \cos^8 \theta \nonumber \\
&&
 -144\, {r}^{7}{a}^{10}{M}^{3} \cos^8 \theta
 -60\, {M}^{2}{r}^{4}{a}^{14}\cos^8 \theta
 +96\, {M}^{3}{r}^{5}{a}^{12}\cos^8 \theta
 -8\, {M}^{3}{r}^{3}{a}^{14} \cos^8 \theta \nonumber \\
&&
 +90\, M{r}^{13}{a}^{6} \cos^6 \theta
 +60\, M{r}^{11}{a}^{8} \cos^6 \theta
 -150\, M{r}^{9}{a}^{10} \cos^6 \theta
 -60\, {M}^{2}{r}^{12}{a}^{6}\cos^6 \theta \nonumber \\
&&
 +180\, {M}^{2}{r}^{10}{a}^{8}\cos^6 \theta
 -120\, M{r}^{7}{a}^{12} \cos^6 \theta
 +120\, {M}^{2}{r}^{8}{a}^{10}\cos^6 \theta
 -120\, {M}^{2}{r}^{6}{a}^{12}\cos^6 \theta \nonumber \\
&&
 +8\, {M}^{3}{r}^{11}{a}^{6}\cos^6 \theta
 -96\, {M}^{3}{r}^{9}{a}^{8}\cos^6 \theta
 +144\,{r}^{7}{a}^{10}{M}^{3} \cos^6 \theta
 -32\, {M}^{3}{r}^{5}{a}^{12}\cos^6 \theta \nonumber \\
&&
 +36\, M{r}^{15}{a}^{4} \cos^4 \theta
 -18\, M{r}^{13}{a}^{6} \cos^4 \theta
 -144\, M{r}^{11}{a}^{8} \cos^4 \theta
 -90\, M{r}^{9}{a}^{10} \cos^4 \theta \nonumber \\
&&
 -12\, {M}^{2}{r}^{14}{a}^{4} \cos^4 \theta
 +108\, {M}^{2}{r}^{12}{a}^{6} \cos^4 \theta
 -120\, {M}^{2}{r}^{8}{a}^{10} \cos^4 \theta
 -24\, {M}^{3}{r}^{11}{a}^{6} \cos^4 \theta \nonumber \\
&&
 +96\, {M}^{3}{r}^{9}{a}^{8} \cos^4 \theta
 -48\, {M}^{3}{r}^{7}{a}^{10} \cos^4 \theta
 +6\, M{r}^{17}{a}^{2} \cos^2 \theta
 -24\, M{r}^{15}{a}^{4} \cos^2 \theta \nonumber \\
&&
 -66\, M{r}^{13}{a}^{6} \cos^2 \theta
 -36\, M{r}^{11}{a}^{8} \cos^2 \theta
 +24\, {M}^{2}{r}^{14}{a}^{4} \cos^2 \theta
 -36\, {M}^{2}{r}^{12}{a}^{6} \cos^2 \theta \nonumber \\
&&
 -60\, {M}^{2}{r}^{10}{a}^{8} \cos^2 \theta
 +24\, {M}^{3}{r}^{11}{a}^{6} \cos^2 \theta
  -32\, {M}^{3}{r}^{9}{a}^{8} \cos^2 \theta,
\end{eqnarray}
\begin{eqnarray}
N_{\theta\phi} &=& 21\,{r}^{13} +12\,{M}^{2}{r}^{7}{a}^{4} \cos^4 \theta
+7\,M{r}^{2}{a}^{10} \cos^4 \theta -113\,M{r}^{4}{a}^{8} \cos^4
\theta
-27\,M{r}^{6}{a}^{6} \cos^4 \theta \nonumber \\
&&
+117\,M{r}^{8}{a}^{4} \cos^4 \theta
-96\,{M}^{2}{r}^{5}{a}^{6} \cos^2 \theta
-24\,{M}^{2}{r}^{7}{a}^{4} \cos^2 \theta
+11\,M{r}^{4}{a}^{8} \cos^2 \theta  \nonumber \\
&&
-99\,M{r}^{6}{a}^{6} \cos^2 \theta
-147\,M{r}^{8}{a}^{4} \cos^2 \theta
-21\,M{r}^{10}{a}^{2} \cos^2 \theta
-22\,M{r}^{2}{a}^{10} \cos^8 \theta \nonumber \\
&&
-17\,M{r}^{4}{a}^{8} \cos^8 \theta
+20\,{M}^{2}{r}^{3}{a}^{8} \cos^8 \theta
-40\,{M}^{2}{r}^{3}{a}^{8} \cos^6 \theta
-96\,{M}^{2}{r}^{5}{a}^{6} \cos^6 \theta \nonumber \\
&&
+15\,M{r}^{2}{a}^{10} \cos^6 \theta
+119\,M{r}^{4}{a}^{8} \cos^6 \theta
+121\,M{r}^{6}{a}^{6} \cos^6 \theta
+20\,{M}^{2}{r}^{3}{a}^{8} \cos^4 \theta \nonumber \\
&&
+192\,{M}^{2}{r}^{5}{a}^{6} \cos^4 \theta
+8\,{r}^{3}{a}^{10} \cos^8 \theta
+4\,{r}^{5}{a}^{8} \cos^8 \theta
-M{a}^{12} \cos^8 \theta \nonumber \\
&&
+M{a}^{12} \cos^6 \theta
+3\,r{a}^{12} \cos^6 \theta
-21\,{r}^{3}{a}^{10} \cos^6 \theta
-71\,{r}^{5}{a}^{8} \cos^6 \theta \nonumber \\
&&
-31\,{r}^{7}{a}^{6} \cos^6 \theta
+9\,{r}^{3}{a}^{10} \cos^4 \theta
-47\,{r}^{5}{a}^{8} \cos^4 \theta
-133\,{r}^{7}{a}^{6} \cos^4 \theta \nonumber \\
&&
-53\,{r}^{9}{a}^{4} \cos^4 \theta
+9\,{r}^{5}{a}^{8} \cos^2 \theta
+{r}^{7}{a}^{6} \cos^2 \theta
-21\,{r}^{9}{a}^{4} \cos^2 \theta \nonumber \\
&&
+12\,{M}^{2}{r}^{7}{a}^{4}
+21\,M{r}^{10}{a}^{2}
+30\,M{r}^{8}{a}^{4}
+5\,M{r}^{6}{a}^{6}
+3\,{r}^{7}{a}^{6}
+3\,{r}^{11}{a}^{2} \cos^2 \theta
\nonumber \\
&&
+8\,r{a}^{12} \cos^8 \theta
+33\,{r}^{11}{a}^{2}
+19\,{r}^{9}{a}^{4} ,
\end{eqnarray}
\begin{eqnarray}
D_{\theta\phi} &=& -{r}^{20}  -3\,{r}^{18}{a}^{2} -3\,{r}^{16}{a}^{4}
-{r}^{14}{a}^{6} - {a}^{20} \cos^{14} \theta -7\,{r}^{18}{a}^{2}
\cos^2 \theta
-105\,{r}^{8}{a}^{12} \cos^8 \theta \nonumber \\
&&
-35\,{r}^{14}{a}^{6} \cos^6 \theta
-21\,{r}^{16}{a}^{4} \cos^4 \theta
-63\,{r}^{14}{a}^{6} \cos^4 \theta
-21\,{r}^{10}{a}^{10} \cos^4 \theta  \nonumber \\
&&
-105\,{r}^{12}{a}^{8} \cos^6 \theta
-35\,{r}^{8}{a}^{12} \cos^6 \theta
-105\,{r}^{10}{a}^{10} \cos^8 \theta
-35\,{r}^{6}{a}^{14} \cos^8 \theta  \nonumber \\
&&
-35\,{r}^{12}{a}^{8} \cos^8 \theta
-63\,{r}^{6}{a}^{14} \cos^{10} \theta
-21\,{r}^{10}{a}^{10} \cos^{10} \theta
-21\,{r}^{14}{a}^{6} \cos^2 \theta \nonumber \\
&&
-63\,{r}^{12}{a}^{8} \cos^4 \theta
-105\,{r}^{10}{a}^{10} \cos^6 \theta
-63\,{r}^{8}{a}^{12} \cos^{10} \theta
-21\,{r}^{4}{a}^{16} \cos^{10} \theta  \nonumber \\
&&
-21\,{r}^{16}{a}^{4} \cos^2 \theta
-7\,{r}^{12}{a}^{8} \cos^2 \theta
-6\,M{r}^{17}{a}^{2}
-12\,M{r}^{15}{a}^{4}
-6\,M{r}^{13}{a}^{6} \nonumber \\
&&
-12\,{M}^{2}{r}^{14}{a}^{4}
-12\,{M}^{2}{r}^{12}{a}^{6}
-8\,{M}^{3}{r}^{11}{a}^{6}
- {r}^{6}{a}^{14} \cos^{14} \theta
-3\, {r}^{2}{a}^{18} \cos^{14} \theta \nonumber \\
&&
-3\, {r}^{4}{a}^{16} \cos^{14} \theta
-7\, {r}^{8}{a}^{12} \cos^{12} \theta
-21\, {r}^{6}{a}^{14} \cos^{12} \theta
-21\, {r}^{4}{a}^{16} \cos^{12} \theta
-7\, {r}^{2}{a}^{18} \cos^{12} \theta \nonumber \\
&&
+6\, M{r}^{5}{a}^{14} \cos^{14} \theta
+8\, {M}^{3}{r}^{3}{a}^{14} \cos^{14} \theta
+12\, M{r}^{3}{a}^{16} \cos^{14} \theta
-12\, {M}^{2}{r}^{2}{a}^{16} \cos^{14} \theta \nonumber \\
&&
+6\, Mr{a}^{18} \cos^{14} \theta
-12\, {M}^{2}{r}^{4}{a}^{14} \cos^{14} \theta
+36\, M{r}^{7}{a}^{12} \cos^{12} \theta
+66\, M{r}^{5}{a}^{14} \cos^{12} \theta   \nonumber \\
&&
-60\, {M}^{2}{r}^{6}{a}^{12} \cos^{12} \theta
+24\, M{r}^{3}{a}^{16} \cos^{12} \theta
-36\, {M}^{2}{r}^{4}{a}^{14} \cos^{12} \theta
+32\, {M}^{3}{r}^{5}{a}^{12} \cos^{12} \theta \nonumber \\
&&
-6\, Mr{a}^{18} \cos^{12} \theta
+24\, {M}^{2}{r}^{2}{a}^{16} \cos^{12} \theta
-24\, {M}^{3}{r}^{3}{a}^{14} \cos^{12} \theta
+90\, M{r}^{9}{a}^{10} \cos^{10} \theta \nonumber \\
&&
+144\, M{r}^{7}{a}^{12} \cos^{10} \theta
-120\, {M}^{2}{r}^{8}{a}^{10} \cos^{10} \theta
+18\, M{r}^{5}{a}^{14} \cos^{10} \theta
+48\, {M}^{3}{r}^{7}{a}^{10} \cos^{10} \theta \nonumber \\
&&
-36\, M{r}^{3}{a}^{16} \cos^{10} \theta
+108\, {M}^{2}{r}^{4}{a}^{14} \cos^{10} \theta
-96\, {M}^{3}{r}^{5}{a}^{12} \cos^{10} \theta
-12\, {M}^{2}{r}^{2}{a}^{16} \cos^{10} \theta \nonumber \\
&&
+24\, {M}^{3}{r}^{3}{a}^{14} \cos^{10} \theta
+120\, M{r}^{11}{a}^{8} \cos^{8} \theta
+150\, M{r}^{9}{a}^{10} \cos^{8} \theta
-120\, {M}^{2}{r}^{10}{a}^{8} \cos^{8} \theta \nonumber \\
&&
-60\, M{r}^{7}{a}^{12} \cos^{8} \theta
+120\, {M}^{2}{r}^{8}{a}^{10} \cos^{8} \theta
-90\, M{r}^{5}{a}^{14} \cos^{8} \theta
+180\, {M}^{2}{r}^{6}{a}^{12} \cos^{8} \theta \nonumber \\
&&
+32\, {M}^{3}{r}^{9}{a}^{8} \cos^{8} \theta
-144\, {M}^{3}{r}^{7}{a}^{10} \cos^{8} \theta
-60\, {M}^{2}{r}^{4}{a}^{14} \cos^{8} \theta
+96\, {M}^{3}{r}^{5}{a}^{12} \cos^{8} \theta \nonumber \\
&&
-8\, {M}^{3}{r}^{3}{a}^{14} \cos^{8} \theta
+90\, M{r}^{13}{a}^{6} \cos^{6} \theta
+60\, M{r}^{11}{a}^{8} \cos^{6} \theta
-150\, M{r}^{9}{a}^{10} \cos^{6} \theta \nonumber \\
&&
-60\, {M}^{2}{r}^{12}{a}^{6} \cos^{6} \theta
+180\, {M}^{2}{r}^{10}{a}^{8} \cos^{6} \theta
-120\, M{r}^{7}{a}^{12} \cos^{6} \theta
+120\, {M}^{2}{r}^{8}{a}^{10} \cos^{6} \theta \nonumber \\
&&
-120\, {M}^{2}{r}^{6}{a}^{12} \cos^{6} \theta
+8\, {M}^{3}{r}^{11}{a}^{6} \cos^{6} \theta
-96\, {M}^{3}{r}^{9}{a}^{8} \cos^{6} \theta
+144\, {M}^{3}{r}^{7}{a}^{10} \cos^{6} \theta \nonumber \\
&&
-32\, {M}^{3}{r}^{5}{a}^{12} \cos^{6} \theta
+36\, M{r}^{15}{a}^{4} \cos^{4} \theta
-18\, M{r}^{13}{a}^{6} \cos^{4} \theta
-144\, M{r}^{11}{a}^{8} \cos^{4} \theta \nonumber \\
&&
-90\, M{r}^{9}{a}^{10} \cos^{4} \theta
-12\, {M}^{2}{r}^{14}{a}^{4} \cos^{4} \theta
+108\, {M}^{2}{r}^{12}{a}^{6} \cos^{4} \theta
-120\, {M}^{2}{r}^{8}{a}^{10} \cos^{4} \theta \nonumber \\
&&
-24\, {M}^{3}{r}^{11}{a}^{6} \cos^{4} \theta
+96\, {M}^{3}{r}^{9}{a}^{8} \cos^{4} \theta
-48\, {M}^{3}{r}^{7}{a}^{10} \cos^{4} \theta
+6\, M{r}^{17}{a}^{2} \cos^{2} \theta \nonumber \\
&&
-24\, M{r}^{15}{a}^{4} \cos^{2} \theta
-66\, M{r}^{13}{a}^{6} \cos^{2} \theta
-36\, M{r}^{11}{a}^{8} \cos^{2} \theta
+24\, {M}^{2}{r}^{14}{a}^{4} \cos^{2} \theta \nonumber \\
&&
-36\, {M}^{2}{r}^{12}{a}^{6} \cos^{2} \theta
-60\, {M}^{2}{r}^{10}{a}^{8} \cos^{2} \theta
+24\, {M}^{3}{r}^{11}{a}^{6} \cos^{2} \theta \nonumber \\
&& -32\, {M}^{3}{r}^{9}{a}^{8} \cos^{2} \theta.
\end{eqnarray}

\section{Curvature square terms}
\label{apendixC} The explicit form of $R^2$,
$R^2_{ij}=R_{ij}R^{ij}$, and $C^2_{ij}=C_{ij}C^{ij}$ is calculated
upto $a^4$-order.

 $R^2$ is given by
\begin{eqnarray}
R^2 &=& - \frac{18 M^2 \sin^2 \theta}{r^6} a^2 \nonumber \\
&& +\frac{6M^2 \sin^2 \theta}{r^8}
 \left ( 4 + 13 \cos^2 \theta + \frac{12 M \sin^2 \theta}{r}\right )a^4 + {\cal O}(a^6).
\label{r2_series}
\end{eqnarray}
$R_{ij}R^{ij}$ takes the form
\begin{eqnarray}
R_{ij}R^{ij} &=& \frac{6M^2}{r^6}
+ \frac{18M^2}{r^8} \left ( 1 -5 \cos^2 \theta - \frac{3M\sin \theta}{r}\right ) a^2 \nonumber \\
&&
+\frac{6M^2}{r^{10}} \left [
\frac{81 M^2 \sin^4 \theta}{r^2}
+\frac{M \sin^2 \theta}{r} \left ( 82 \cos^2 \theta - 17\right )
+ 81 \cos^4 \theta - 18 \cos^2 \theta - 17
\right ] a^4 \nonumber \\
&& + {\cal O}(a^6).
\label{rij2_series}
\end{eqnarray}
Finally, $C_{ij}C^{ij}$ is given by
\begin{equation}
C_{ij}C^{ij} = \frac{3528 M^4 \sin^2 \theta}{r^{16}}
\left ( 1 - \frac{2M \sin^2 \theta}{r}\right ) a^4 + {\cal O}(a^6).
\label{cij2_series}
\end{equation}

\section{$E^{(k)}_{ij}$ up to $a^4$-order}
\label{apendixD} We show  the explicit form  $E_{ij}^{(k)}$.  All
other terms not listed here are zero.

 $E_{ij}^\1$ is given by
\begin{eqnarray}
E_{rr}^\1 &=& \frac{9M^2\sin^2 \theta}{r^6 \sqrt{1-\frac{2M}{r}}} a^2 \nonumber \\
&&
-\frac{3M^2 \sin^2\theta}{r^{8}\left( 1-\frac{2M}{r}\right )^{3/2}}
  \left [ 7 + 10 \cos^2 \theta
  +\frac{M}{r} \left ( 4 -35 \cos^2 \theta \right )
   - \frac{30M^2 \sin^2\theta}{r^2} \right] a^4 \nonumber \\
&&
  + {\cal O}(a^6),
\label{E1_rr}
\end{eqnarray}
\begin{eqnarray}
E_{r\theta}^\1 &=& -\frac{12M^2\sin^3 \theta \cos \theta}{r^7}
     \sqrt{ 1-\frac{2M}{r} } a^4
  + {\cal O}(a^6),
\label{E1_rth}
\end{eqnarray}
\begin{eqnarray}
E_{\theta\theta}^\1 &=& -\frac{9M^2\sin^2 \theta}{r^4} \sqrt{1-\frac{2M}{r}} a^2 \nonumber \\
&&
+\frac{3M^2 \sin^2\theta}{r^{6} \sqrt{1-\frac{2M}{r}}}
  \left [ 4 + 10 \cos^2 \theta
  +\frac{M}{r} \left ( 4 -35 \cos^2 \theta \right )
  - \frac{30M^2 \sin^2\theta}{r^2} \right] a^4 \nonumber \\
&&
  + {\cal O}(a^6),
\label{E1_thth}
\end{eqnarray}
\begin{eqnarray}
E_{\phi\phi}^\1 &=& -\frac{27M^2\sin^4 \theta}{r^4} \sqrt{1-\frac{2M}{r}} a^2 \nonumber \\
&&
+\frac{9M^2 \sin^4\theta}{r^{6} \sqrt{1-\frac{2M}{r}}}
  \left [ 1 + 13 \cos^2 \theta
  + \frac{M}{r} \left ( 4 - 35 \cos^2 \theta \right )
  - \frac{30M^2 \sin^2\theta}{r^2} \right] a^4 \nonumber \\
&&
  + {\cal O}(a^6).
\label{E1_phph}
\end{eqnarray}
Explicitly, we show that $ E^{(2)}_{ij}=0$.

\noindent $E_{ij}^\3$ is given by
\begin{eqnarray}
E_{rr}^\3 &=& -\frac{9M^2\sin^2 \theta}{r^6 \sqrt{1-\frac{2M}{r}}} a^2 \nonumber \\
&&
+\frac{3M^2 \sin^2\theta}{r^{8}\left( 1-\frac{2M}{r}\right )^{3/2}}
  \left [ 7 + 10 \cos^2 \theta
  + \frac{M}{r} \left ( 4 - 35 \cos^2 \theta \right )
  - \frac{30M^2 \sin^2\theta}{r^2} \right] a^4 \nonumber \\
&&
  + {\cal O}(a^6),
\label{E3_rr}
\end{eqnarray}

\begin{eqnarray}
E_{r\theta}^\3 &=&\frac{12M^2\sin^3 \theta \cos \theta}{r^7}
     \sqrt{ 1-\frac{2M}{r} } a^4 \nonumber \\
  &+& {\cal O}(a^6),
\label{E3_rth}
\end{eqnarray}
\begin{eqnarray}
E_{\theta\theta}^\3 &=& \frac{9M^2\sin^2 \theta}{r^4} \sqrt{1-\frac{2M}{r}} a^2 \nonumber \\
&&
-\frac{3M^2 \sin^2\theta}{r^{6} \sqrt{1-\frac{2M}{r}}}
  \left [ 4 +  10 \cos^2 \theta
  + \frac{M}{r} \left ( 4 - 35 \cos^2 \theta \right )
  - \frac{30M^2 \sin^2\theta}{r^2} \right] a^4 \nonumber \\
&&
  + {\cal O}(a^6),
\label{E3_thth}
\end{eqnarray}
\begin{eqnarray}
E_{\phi\phi}^\3 &=& \frac{27M^2\sin^4 \theta}{r^4} \sqrt{1-\frac{2M}{r}} a^2 \nonumber \\
&&
-\frac{9M^2 \sin^4\theta}{r^{6} \sqrt{1-\frac{2M}{r}}}
  \left [ 1 + 13 \cos^2 \theta
  + \frac{M}{r} \left ( 4 - 35 \cos^2 \theta \right )
  - \frac{30M^2 \sin^2\theta}{r^2} \right] a^4 \nonumber \\
&&
  + {\cal O}(a^6).
\label{E3_phph}
\end{eqnarray}

$E_{ij}^\4$ is
\begin{eqnarray}
E_{rr}^\4 &=& -\frac{72M^2}{r^8 \sqrt{1-\frac{2M}{r}}}
\left ( 7 - 9 \cos^2 \theta - \frac{12M \sin^2 \theta}{r}\right )a^2 \nonumber \\
&&
+\frac{6M^2 }{r^{10}\left( 1-\frac{2M}{r}\right )^{3/2}}
  \Big [ 140  + 684 \cos^2 \theta -984 \cos^4 \theta  \nonumber \\
&&
 + \frac{M}{r} \left ( 52 - 3780 \cos^2 \theta +4024 \cos^4 \theta \right ) \nonumber \\
&& \left .
  - \frac{M^2 }{r^2} \left ( 1891 - 7254 \cos^2 \theta +5363 \cos^4 \theta \right )
  + \frac{2502 M^3 \sin^4 \theta }{r^3}
\right] a^4 \nonumber \\
&&
  + {\cal O}(a^6),
\label{E4_rr}
\end{eqnarray}
\begin{eqnarray}
E_{r\theta}^\4 &=& \frac{72M^2\sin \theta \cos \theta}{r^7
     \sqrt{ 1-\frac{2M}{r}} } \left ( 7 - \frac{15M}{r} \right ) a^2 \nonumber \\
&&
 + \frac{24 M^2 \sin \theta \cos\theta}{r^9 \left ( 1-\frac{2M}{r}\right )^{3/2}}
    \Big [ 90 - 246 \cos^2 \theta
  -\frac{M}{r} \left ( 627 -1298 \cos^2 \theta \right ) \nonumber \\
&& \left .
  +\frac{M^2}{r^2} \left ( 1500 -2215 \cos^2 \theta \right )
  -\frac{1206M^3\sin^2 \theta}{r^3} \right ] a^4
  + {\cal O}(a^6),
\label{E4_rth}
\end{eqnarray}
\begin{eqnarray}
E_{\theta\theta}^\4 &=& \frac{72M^2}{r^6  \sqrt{1-\frac{2M}{r}}}
  \left [ 18 -17 \cos^2 \theta
   -\frac{M}{r} \left ( 81 - 79 \cos^2 \theta \right )
   + \frac{90M^2 \sin^2 \theta}{r^2}\right ]  a^2 \nonumber \\
&&
-\frac{6M^2 }{r^{8} \sqrt{1-\frac{2M}{r}}}
  \Big [ 224 + 1452 \cos^2 \theta -1608 \cos^4 \theta \nonumber \\
&&
  +\frac{M}{r} \left ( 316 -9888 \cos^2 \theta + 9424 \cos^4 \theta\right )
  - \frac{M^2}{r^2} \left ( 6793 -24582 \cos^2 \theta +17789 \cos^4 \theta \right ) \nonumber \\
&& \left .
  + \frac{M^3}{r^3} \left ( 10746 -21492\cos^2 \theta +10746 \cos^4 \theta \right )
  \right] a^4
  + {\cal O}(a^6),
\label{E4_thth}
\end{eqnarray}
\begin{eqnarray}
E_{\phi\phi}^\4 &=& \frac{72M^2\sin^2 \theta}{r^6} \sqrt{1-\frac{2M}{r}}
 \left ( 17 -16\cos^2 \theta -\frac{45M}{r} \sin^2 \theta \right ) a^2 \nonumber \\
&&
-\frac{6M^2 \sin^2\theta}{r^{8} \sqrt{1-\frac{2M}{r}}}
  \Big [ 92 + 1416 \cos^2 \theta - 1440 \cos^4 \theta \nonumber \\
&&
  + \frac{M}{r} \left ( 652 - 9708 \cos^2 \theta \right )
  - \frac{M^2}{r^2} \left ( 6157 -22830 \cos^2 \theta +16678\cos^4 \theta \right ) \nonumber \\
&& \left .
  + \frac{M^3}{r^3} \left ( 9234 -18468\cos^2 \theta +9234 \cos^4 \theta \right )
  \right] a^4
  + {\cal O}(a^6).
\label{E4_phph}
\end{eqnarray}

$E_{ij}^\5$ is given by
\begin{eqnarray}
E_{rr}^\5 &=&  -\frac{4\mu W^2 M^2}{r^6 \sqrt{1-\frac{2M}{r}}} \nonumber \\
&&
 -\frac{\mu W^2 M^2 }
          {2r^{8}\left( 1-\frac{2M}{r}\right )^{3/2}}
  \Big [ 168  - 229 \cos^2 \theta  \nonumber \\
&& \left .
 - \frac{M}{r} \left ( 786 - 910 \cos^2 \theta \right )
   + \frac{904 M^2 \sin^2 \theta}{r^2} \right ] a^2 \nonumber \\
&&
  + \frac{\mu W^2 M^2 }{2r^{10} \left( 1-\frac{2M}{r} \right )^{5/2}}
  \Big [ 364 + 662 \cos^2 \theta - 1338 \cos^4 \theta \nonumber \\
&&
  - \frac{M}{r} \left ( 1026 + 8021 \cos^2 \theta -10221 \cos^4 \theta \right ) \nonumber \\
&&
  - \frac{M^2}{r^2} \left ( 4300 -31809 \cos^2 \theta +28710 \cos^4 \theta \right ) \nonumber \\
&&
  + \frac{M^3}{r^3} \left ( 17532 -52536 \cos^2 \theta +35004 \cos^4 \theta \right ) \nonumber \\
&& \left .
  - \frac{M^4}{r^4} \left ( 15228 -31056 \cos^2 \theta +15228 \cos^4 \theta \right )
\right] a^4
  + {\cal O}(a^6),
\label{E5_rr}
\end{eqnarray}
\begin{eqnarray}
E_{r\theta}^\5 &=& \frac{3\mu W^2}{2} \frac{M^2 \sin\theta \cos\theta}
                   {r^7 \sqrt{1-\frac{2M}{r}}}
\left ( 8 - \frac{11M}{r} \right )a^2 \nonumber \\
&&
+\frac{3\mu W^2}{2} \frac{M^2 \sin\theta \cos\theta}
          {r^{9}\left( 1-\frac{2M}{r}\right )^{3/2}}
  \Big [ 130  - 178 \cos^2 \theta
  -\frac{M}{r} \left ( 949 -1103 \cos^2 \theta \right ) \nonumber \\
&& \left .
  +\frac{M^2}{r^2} \left ( 2194 -2315 \cos^2 \theta \right )
  -\frac{1642M^3\sin^2 \theta}{r^3} \right ] a^4
  + {\cal O}(a^6),
\label{E5_rth}
\end{eqnarray}
\begin{eqnarray}
E_{r\phi}^\5 &=& -\frac{252M^2\sin^2 \theta \cos \theta}{r^8}
      \left ( 4 - \frac{11M}{r} \right ) a^2 \nonumber \\
&&
 - \frac{12 M^2 \sin^2 \theta \cos\theta}{r^{10}}
    \left [ 234 - 990 \cos^2 \theta
  -\frac{M}{r} \left ( 1569 -4068 \cos^2 \theta \right )
  +\frac{2419M^2\sin^2 \theta}{r^2} \right ] a^4 \nonumber \\
&&
  + {\cal O}(a^6),
\label{E5_rph}
\end{eqnarray}
\begin{eqnarray}
E_{\theta\theta}^\5 &=& \frac{7 \mu W^2 M^2}{2r^4}  \sqrt{1-\frac{2M}{r}} \nonumber \\
&& + \frac{ \mu W^2 M^2}{2r^6 \sqrt{1-\frac{2M}{r}}}
  \left [ 240 -248 \cos^2 \theta
   +\frac{1145M\cos^2\theta}{r}
   + \frac{1298M^2 \sin^2 \theta}{r^2}\right ]  a^2 \nonumber \\
&&
-\frac{ \mu W^2 M^2 }{4 r^{8} \left ( 1-\frac{2M}{r}\right )^{3/2}}
  \Big [ 738 + 2232 \cos^2 \theta - 2928 \cos^4 \theta \nonumber \\
&&
  -\frac{M}{r} \left ( 1236 +24240 \cos^2 \theta - 25338 \cos^4 \theta\right )
  - \frac{M^2}{r^2} \left ( 17672 -98136 \cos^2 \theta +80349 \cos^4 \theta \right ) \nonumber \\
&&
  + \frac{M^3}{r^3} \left ( 62664 -173700\cos^2 \theta +111036 \cos^4 \theta \right ) \nonumber \\
&& \left .
  - \frac{M^4}{r^4} \left ( 56532 -113064\cos^2 \theta +56532 \cos^4 \theta \right )
  \right] a^4
  + {\cal O}(a^6),
\label{E5_thth}
\end{eqnarray}
\begin{eqnarray}
E_{\theta\phi}^\5 &=& -\frac{168M^2 \sin^3 \theta}{r^7}
  \left ( 9 - \frac{48M}{r} +\frac{61M^2}{r^2} \right )  a^2 \nonumber \\
&&
 +\frac{12M^2 \sin^3 \theta }{r^{9}}
  \Big [ 24 + 1320 \cos^2 \theta -1608 \cos^4 \theta
  +\frac{M}{r} \left ( 666 -8862 \cos^2 \theta \right ) \nonumber \\
&& \left .
  - \frac{M^2}{r^2} \left ( 5294 - 16858 \cos^2 \theta\right )
  + \frac{8575M^3\sin^2\theta}{r^3}
  \right] a^4
  + {\cal O}(a^6),
\label{E5_thph}
\end{eqnarray}
\begin{eqnarray}
E_{\phi\phi}^\5 &=& \frac{7 \mu W^2 M^2\sin^2 \theta}{2r^4} \sqrt{1-\frac{2M}{r}} \nonumber \\
&&
 + \frac{ \mu W^2 M^2\sin^2 \theta}{2r^6 \sqrt{1-\frac{2M}{r}}}
 \left [ 712 -720\cos^2 \theta
 -\frac{M}{r} \left ( 3426 -3449 \cos^2 \theta \right )
 +\frac{4018M^2 \sin^2 \theta}{r^2} \right ] a^2 \nonumber \\
&&
-\frac{\mu W^2 M^2 \sin^2\theta}{4 r^{8} \left (1-\frac{2M}{r} \right )^{3/2}}
  \Big [ 306 + 10140 \cos^2 \theta - 10404 \cos^4 \theta \nonumber \\
&&
  + \frac{M}{r} \left ( 4788 - 91072 \cos^2 \theta + 86146 \cos^4 \theta \right ) \nonumber \\
&&
  - \frac{M^2}{r^2} \left ( 52752 -303736 \cos^2 \theta +250869\cos^4 \theta \right ) \nonumber \\
&&
  + \frac{M^3}{r^3} \left ( 146184 -448836\cos^2 \theta +302652 \cos^4 \theta \right ) \nonumber \\
&& \left .
  - \frac{M^4}{r^4} \left ( 124532 -249064\cos^2 \theta +124532 \cos^4 \theta \right )
  \right] a^4
  + {\cal O}(a^6).
\label{E5_phph}
\end{eqnarray}

Finally, $E_{ij}^\6$ is given by
\begin{eqnarray}
E_{rr}^\6 &=&
 \frac{5\mu^2 W^4 M^2}{4r^6 \sqrt{1-\frac{2M}{r}}} \nonumber \\
&&
 +\left [  
-\frac{1}{4} \frac{\mu^2 W^4 M^2}{r^8 \left ( 1-\frac{2M}{r} \right)^{3/2}}
 \left \{
  -10 + 70 \cos^2 \theta 
  +\frac{M}{r} \left ( 123 - 238 \cos^2 \theta \right ) \nonumber \right . \right . \\
&& \left . \left .
  -\frac{M^2}{r^2} \left ( 196 \sin^2 \theta \right )
 \right \}
-\frac{1}{4} \frac{M}{r^{11} \sqrt{ 1-\frac{2M}{r} }}
 \left (
  2856 - 8568 \cos^2 \theta
 \right )
\right ] a^2 \nonumber \\
&&
 +\left [ 
\frac{1}{8} \frac{\mu^2 W^4 M^2 }{r^{10} \left ( 1-\frac{2M}{r}\right )^{5/2}}
\left \{ -2 -94 \cos^2 \theta +756 \cos^4 \theta \right . \right . \nonumber \\
&&
  + \frac{M}{r} \left ( 108 + 2422 \cos^2 \theta -5040 \cos^4 \theta \right ) \nonumber \\
&&
  + \frac{M^2}{r^2} \left ( 1166 - 11242 \cos^2 \theta + 12472 \cos^4 \theta \right ) \nonumber \\
&&
  + \frac{M^3}{r^3} \left ( -5484 + 19080 \cos^2 \theta - 13596 \cos^4 \theta \right ) \nonumber \\
&& \left . 
  + \frac{M^4}{r^4} \left ( 5532 - 11064 \cos^2 \theta + 5532 \cos^4 \theta \right ) \right \} \nonumber \\
&& 
+\frac{1}{8} \frac{M^3 }{r^{13} \left ( 1-\frac{2M}{r}\right )^{5/2}}
\left \{ 32592 + 158544 \cos^2 \theta - 331104 \cos^4 \theta \right . \nonumber \\
&&
  + \frac{M}{r} \left ( -261264 - 647136 \cos^2 \theta + 1456848 \cos^4 \theta \right ) \nonumber \\
&&
  + \frac{M^2}{r^2} \left ( 951648 + 79200 \cos^2 \theta - 1567872 \cos^4 \theta \right ) \nonumber \\
&&
  + \frac{M^3}{r^3} \left ( -1691520 + 2306880 \cos^2 \theta - 615360 \cos^4 \theta \right ) \nonumber \\
&& \left . \left .
  + \frac{M^4}{r^4} \left ( 1145088 \sin^4 \theta \right ) \right \}
 \right ] a^4
  + {\cal O}(a^6),
\label{E6_rr}
\end{eqnarray}
\begin{eqnarray}
E_{r\theta}^\6 &=&
 \left [ 
-\frac{9}{2} \frac{\mu^2 W^4 M^2 \cos\theta\sin\theta}{r^7 }
  \sqrt { 1- \frac{2M}{r}} \right . \nonumber \\
&& \left .
+\frac{3}{2}\frac{M^3 \cos\theta\sin\theta}{r^{10}\sqrt{1-\frac{2M}{r}}}
\left ( 2758 - 5642 \frac{M}{r} \right )
 \right  ] a^2 \nonumber \\
&&
 +\left [ 
-\frac{3}{2}\frac{\mu^2 W^4 M^2 \cos\theta\sin\theta}{r^9 \sqrt{1-\frac{2M}{r}}}
\left \{
-28\cos^2\theta + \frac{M}{r}\left ( -35 + 150 \cos^2 \theta \right ) \right . \right . \nonumber \\
&& \left .
  + \frac{M^2}{r^2} \left ( 146 - 264 \cos^2 \theta  \right ) 
  - \frac{M^3}{r^3} \left ( 152 \sin^2 \theta \right ) 
\right \} \nonumber \\
&&
-\frac{3}{2}\frac{M^3 \cos\theta\sin\theta}{r^{12} \sqrt{1-\frac{2M}{r}}}
\left \{
 3106 + 36928 \cos^2\theta 
- \frac{M}{r}\left ( 3026 + 161020 \cos^2 \theta \right ) \right . \nonumber \\
&& \left . 
  - \frac{M^2}{r^2} \left ( 34584 - 202414 \cos^2 \theta  \right )
  + \frac{M^3}{r^3} \left ( 56172 \sin^2 \theta \right ) 
\right \} \nonumber \\
&&
+\frac{3}{2}\frac{M^4 \sin^2\theta}{r^{14} \sqrt{1-\frac{2M}{r}}}
\left \{
 -2352 
 + \frac{M}{r}\left ( 9408 - 4704 \cos^2 \theta \right ) \right . \nonumber \\
&& \left . \left .
  - \frac{M^2}{r^2} \left ( 9408 \sin^2 \theta \right ) 
\right \}
 \right  ] a^4
  + {\cal O}(a^6),
\label{E6_rth}
\end{eqnarray}
\begin{eqnarray}
E_{r\phi}^\6 &=&
 -\frac{9}{4} \frac{\mu W^2 M^2 \cos\theta\sin^2\theta}{r^8}
\left ( 4 + \frac{83M}{r}\right )
  a^2 \nonumber \\
&&
 +\left [
 \frac{3}{4}\frac{\mu W^2 M^2 \cos\theta\sin^2\theta}{r^{10}}
\left \{
 -42 + 90 \cos^2\theta \right . \right . \nonumber \\
&& \left . \left .
 + \frac{M}{r} \left ( 273 + 2656 \cos^2\theta \right ) 
 + \frac{M^2}{r^2} \left ( 1928 \sin^2\theta \right )
\right \}
 \right  ] a^4
  + {\cal O}(a^6),
\label{E6_rph}
\end{eqnarray}
\begin{eqnarray}
E_{\theta\theta}^\6 &=&
-\frac{1}{4}\frac{ \mu^2 W^4 M^2}{r^4}  \sqrt{1-\frac{2M}{r}} \nonumber \\
&&
 +\left [ 
\frac{1}{4}\frac{\mu^2 W^4 M^2}{r^6 \sqrt{1-\frac{2M}{r}}}
\left \{
 3 + 8 \cos^2 \theta 
 - \frac{M}{r} \left ( 3 + 20 \cos^2\theta\right )
 - \frac{M^2}{r^2} \left ( 8 \sin^2\theta \right )
\right \} \right . \nonumber \\
&& 
+\frac{1}{4}\frac{M^3}{r^9 \sqrt{1-\frac{2M}{r}}}
\left \{
 -2520 - 336 \cos^2 \theta 
 + \frac{M}{r} \left ( 17808 - 12096 \cos^2\theta\right ) \right . \nonumber \\
&& \left . \left .
 - \frac{M^2}{r^2} \left ( 25536 \sin^2\theta \right )
\right \} 
 \right  ] a^2 \nonumber \\
&&
 +\left [ 
-\frac{1}{8}\frac{\mu^2 W^4 M^2 }{r^8 \left ( 1-\frac{2M}{r} \right )^{3/2}}
\left \{
-18 -54 \cos^2\theta + 180 \cos^4 \theta \right . \right . \nonumber \\
&&
 + \frac{M}{r} \left ( 114 + 486 \cos^2\theta -1056 \cos^4\theta \right ) 
 - \frac{M^2}{r^2} \left ( 10 + 1830 \cos^2\theta -2319 \cos^4\theta \right ) \nonumber \\
&& \left .
 - \frac{M^3}{r^3} \left ( 804 + 3168 \cos^2\theta + 2364 \cos^4\theta \right ) 
 + \frac{M^4}{r^4} \left ( 1020 \sin^4\theta \right ) 
\right \} \nonumber \\
&&
-\frac{1}{8}\frac{M^3 }{r^{11} \left ( 1-\frac{2M}{r} \right )^{3/2}}
\left \{
 24048 + 38304 \cos^2\theta - 126624 \cos^4 \theta \right . \nonumber \\
&&
 - \frac{M}{r} \left ( 260784 - 340944 \cos^2\theta - 182640 \cos^4\theta \right ) \nonumber \\
&&
 + \frac{M^2}{r^2} \left ( 1156416 - 2631168 \cos^2\theta + 1206240 \cos^4\theta \right ) \nonumber \\
&& \left .
 - \frac{M^3}{r^3} \left ( 2296704 - 5261376 \cos^2\theta + 2964672 \cos^4\theta \right ) 
 + \frac{M^4}{r^4} \left ( 1669248 \sin^4\theta \right ) 
\right \} \nonumber \\
&&
-\frac{1}{8}\frac{M^4 }{r^{14} \left ( 1-\frac{2M}{r} \right )^{3/2}}
\left \{
 -28224 \sin^2\theta  
 + \frac{M}{r} \left ( 112896 - 169344 \cos^2\theta + 56448 \cos^4\theta \right ) \right . \nonumber \\
&& \left . \left .
 - \frac{M^2}{r^2} \left ( 112896 \sin^4\theta \right ) 
\right \}
 \right  ] a^4
  + {\cal O}(a^6),
\label{E6_thth}
\end{eqnarray}
\begin{eqnarray}
E_{\theta\phi}^\6 &=&
 \frac{3}{2}\frac{\mu W^2 M^2 \sin^3 \theta }{r^9 \left ( 1-\frac{2M}{r} \right )}
\left ( 9 - \frac{108M}{r} + \frac{209M^2}{r^2} \right )
 a^2 \nonumber \\
&&
 +\left [ 
 \frac{3}{4}\frac{\mu W^2 M^2 \sin\theta}{r^9 \left ( 1-\frac{2M}{r}\right )}
\left \{
-48 + 168 \cos^2\theta -120 \cos^4 \theta \right . \right . \nonumber \\
&&
+\frac{M}{r} \left ( 443 -2119 \cos^2\theta +1676\cos^4 \theta\right ) \nonumber \\
&& \left .
-\frac{M^2}{r^2} \left ( 2421 -8282 \cos^2\theta +5861\cos^4 \theta\right ) 
+ \frac{M^3}{r^3} \left ( 4574 \sin^4\theta \right )
\right \} \nonumber \\
&& \left .
+ \frac{3}{4} \frac{M^4 \sin^2\theta}{r^{14}\sqrt{1-\frac{2M}{r}}}
\left ( 4704 - 9408 \sin^2\theta \frac{M}{r}\right )
 \right  ] a^4
  + {\cal O}(a^6),
\label{E6_thph}
\end{eqnarray}
\begin{eqnarray}
E_{\phi\phi}^\6 &=&
-\frac{1}{4}\frac{ \mu^2 W^4 M^2 \sin^2\theta}{r^4}  \sqrt{1-\frac{2M}{r}} \nonumber \\
&&
 +\left [ 
-\frac{1}{4}\frac{\mu^2 W^4 M^2 \sin^2\theta}{r^6 \sqrt{1-\frac{2M}{r}}}
\left \{
 10 - 21 \cos^2 \theta 
 - \frac{M}{r} \left ( 81 - 104 \cos^2\theta\right )
 + \frac{M^2}{r^2} \left ( 124 \sin^2\theta \right )
\right \} \right . \nonumber \\
&& 
-\frac{1}{4}\frac{M^3 \sin^2\theta}{r^9 \sqrt{1-\frac{2M}{r}}}
\left \{
 -5376 + 8232 \cos^2 \theta 
 + \frac{M}{r} \left ( 23520 - 29232 \cos^2\theta\right ) \right . \nonumber \\
&& \left . \left .
 - \frac{M^2}{r^2} \left ( 25536 \sin^2\theta \right )
\right \} 
 \right  ] a^2 \nonumber \\
&&
 +\left [ 
\frac{1}{8}\frac{\mu^2 W^4 M^2 \sin^2\theta}{r^8 \left ( 1-\frac{2M}{r} \right )^{3/2}}
\left \{
-36 + 150 \cos^2\theta - 222 \cos^4 \theta \right . \right . \nonumber \\
&&
 + \frac{M}{r} \left ( 240 -1864 \cos^2\theta + 2080 \cos^4\theta \right ) 
 - \frac{M^2}{r^2} \left ( 762 - 6262 \cos^2\theta + 5979 \cos^4\theta \right ) \nonumber \\
&& \left .
 + \frac{M^3}{r^3} \left ( 1500 -7560 \cos^2\theta + 6060 \cos^4\theta \right ) 
 - \frac{M^4}{r^4} \left ( 1292 \sin^4\theta \right ) 
\right \} \nonumber \\
&&
+\frac{1}{8}\frac{M^3 \sin^2\theta}{r^{11} \left ( 1-\frac{2M}{r} \right )^{3/2}}
\left \{
 7920 - 164592 \cos^2\theta + 220944 \cos^4 \theta \right . \nonumber \\
&&
 - \frac{M}{r} \left ( 72096 - 1141968 \cos^2\theta + 1332672 \cos^4\theta \right ) \nonumber \\
&&
 + \frac{M^2}{r^2} \left ( 297504 - 2816544 \cos^2\theta + 2787552 \cos^4\theta \right ) \nonumber \\
&& \left .
 - \frac{M^3}{r^3} \left ( 586368 - 2814720 \cos^2\theta + 2228352 \cos^4\theta \right ) 
 + \frac{M^4}{r^4} \left ( 432768 \sin^4\theta \right ) 
\right \} \nonumber \\
&&
+\frac{1}{8}\frac{M^4 \sin^2\theta}{r^{14} \left ( 1-\frac{2M}{r} \right )^{3/2}}
\left \{
 28224   
 - \frac{M}{r} \left ( 112896 +56448 \cos^2\theta \right ) \right . \nonumber \\
&& \left . \left .
 + \frac{M^2}{r^2} \left ( 112896 \sin^2\theta \right ) 
\right \}
 \right  ] a^4
  + {\cal O}(a^6),
\label{E6_phph}
\end{eqnarray}

\bibliographystyle{apsrev}

\end{document}